\documentstyle[12pt]{article}

\font\uno=cmcsc10 scaled 1200
\font\dos=cmti10 scaled 1200

\def\refname{\uno References} 
\def\thebibliography#1{\section*{\refname}
\list 
 {\arabic{enumi}.}{\settowidth\labelwidth{[#1]}\leftmargin\labelwidth 
 \advance\leftmargin\labelsep 
 \usecounter{enumi}} 
 \def\newblock{\hskip .11em plus .33em minus .07em} 
 \sloppy\clubpenalty4000\widowpenalty4000 
 \sfcode`\.=1000\relax} 
 
\begin{document}
\title{{Nonsupersymmetric gauge coupling unfication in 
[SU(6)]$^4\times$ Z$_4$ and proton stability }}
\author{{\uno A. P\'erez-Lorenzana$^1$\thanks{e-mail: abdel@fis.cinvestav.mx},
William A. Ponce$^{1,2}$\thanks{e-mail: wponce@fis.cinvestav,mx}, and 
Arnulfo Zepeda$^1$}\\
{\dos 1-Departamento de F\'{\i}sica,} \\
{\dos Centro de Investigaci\'on y de  Estudios Avanzados del IPN.}\\ 
{\dos Apdo. Postal 14-740, 07000, M\'exico D. F., M\'exico.}\\
{\dos 2-Departamento de F\'\i sica, Universidad de Antioquia} \\
{\dos A.A. 1226, Medell\'\i n, Colombia.}}
\date{}
\maketitle
{{\uno Abstract}.\small
We systematically study the three family extension of the Pati-Salam 
gauge group with an anomaly-free single irreducible representation which 
contains the known quarks and leptons without mirror fermions. 
In the context of this model we implement the survival 
hypothesis, the modified horizontal survival hypothesis, and
calculate  the tree level masses for the gauge boson and fermion fields. We
also use the extended survival hypothesis in order to 
calculate the  mass scales using the
renormalization group equation. The interacting Lagrangean 
with all the known and predicted gauge interactions is explicitly displayed. 
Finally the stability of the proton in this model is established.
\\[2ex]
}


\section{\uno Introduction}
The renormalizability of the original Pati-Salam\cite{ps} model for 
unification of flavors and forces rests on the existence of conjugate or 
mirror partners of ordinary fermions. Mirror fermions are fermions with 
quantum numbers with respect to the Standard Model (SM) gauge group 
SU(3)$_C\otimes$SU(2)$_L\otimes$U(1)$_Y$ 
identical to those of the known quarks and leptons, except that they 
have opposite handedness from ordinary fermions. Their existence vitiate 
the survival hypothesis~\cite{sh} according to which chiral fermions that 
can pair off while respecting a symmetry will do so, acquiring 
masses grater than or equal to the mass scale of that symmetry.

Today we know how to cancel anomalies without 
introducing unwanted mirror fer\-mions. As a matter of fact, the 
three family extension of the Pati-Salam model without mirror fermions 
was presented recently in the literature, with some aspects of 
the model briefly analyzed in the original reference\cite{pz}. 
But a systematic analysis of this model is still 
lacking. In what follows we do such analysis, paying special 
attention to the implementation of the survival hypothesis~\cite{sh} and of
the modified horizontal survival  hypothesis~\cite{hsh}. (For a technical
explanation of the terminology used in this  article see Appendix A.)

The model under consideration unifies non-gravitational 
forces with three families of flavors, using the gauge group 

\[G\equiv 
SU(6)_L\otimes SU(6)_R\otimes SU(6)_{CR}\otimes SU(6)_{CL}\times Z_4 \]

\noindent
where $\otimes$ indicates a direct product, $\times$ a semidirect one, 
and Z$_4\equiv$(1,P,P$^2$,P$^3$) is the four-element cyclic group acting 
upon [SU(6)]$^4$ such that if (A,B,C,D) is a representation of 
[SU(6)]$^4$ with A a representation of the first factor, B of the second, 
C of the third, and D of the fourth, then P(A,B,C,D)=(B,C,D,A) and then 
Z$_4$(A,B,C,D)$\equiv $(A,B,C,D) $\oplus$ (B,C,D,A) $\oplus$ (C,D,A,B) 
$\oplus$ (D,A,B,C). The electric charge operator in $G$ is defined 
as\cite{pz}
\begin{equation}
Q_{EM}=T_{ZL}+T_{ZR}+[Y_{(B-L)_L}+Y_{(B-L)_R}]/2,
\end{equation}
\noindent
where $(B-L)_{L(R)}$ stands for the local Abelian factor of 
$(Baryon-Lepton)_{L(R)}$ 
hypercharge associated with the diagonal generators 
$Y_{(B-L)_{L(R)}}=\!
Diag(\frac{1}{3},\frac{1}{3},\frac{1}{3},-1,1,-1)_{L(R)}$ of SU(6)$_{CL(CR)}$

The irreducible representation (irrep) of $G$ which contains the known 
fermions is 

\[\psi(144)=Z_4\psi(\bar{6},1,1,6)=\psi(\bar{6},1,1,6)\oplus
\psi(1,1,6,\bar{6})\oplus\psi(1,6,\bar{6},1)\oplus\psi(6,\bar{6},1,1).\]

The model described by the structure 
$[G,\psi(144)]$ is a grand unification model which contains
the three family SM gauge group, the three family left-right 
symmetric extension of the SM\cite{moha} 
[SU(3)$_C\otimes$SU(2)$_L\otimes$SU(2)$_R\otimes$U(1)$_{(B-L)}$] and 
the three family chiral color extension of the SM\cite{gf} 
[SU(3)$_{CR}\otimes$SU(3)$_{CL}\otimes$SU(2)$_L$ $\otimes$U(1)$_Y$].
Finally, $[G,\psi(144)]$ is the chiral extension of the 
vector-color-like model described by~\cite{mex,ponce} 
$G^V\equiv $ SU(6)$_L\otimes$ SU(6)$_C\otimes$ SU(6)$_R\times$Z$_3$ and 
$\psi^V(108)
=\psi^V(\bar{6},6,1)\oplus\psi^V(6,1,\bar{6})\oplus\psi^V(1,\bar{6},6)$, 
where SU(6)$_C$ in $G^V$ is the diagonal subgroup of 
SU(6)$_{CR}\otimes$ SU(6)$_{CL}\subset G$, and $\psi^V(108)\subset \psi(144)$

That [$G,\psi(144)]$ is free of anomalies and does  not
contain mirror fermions follows from its particle content. To see this we 
first show that there is a unique way
to embed the SM gauge group for three families in [G,$\psi (144)$]\cite{pz} and then   write
the quantum numbers for $\psi(144)$ with respect to the subgroups of the  SM
which are [the notation designates behavior under (SU(3)$_C$, SU(2)$_L$,
U(1)$_Y$)]:

\noindent
$\psi(\bar{6},1,1,6)\sim 3(3,2,1/3)\oplus 6(1,2,-1)\oplus3(1,2,1)$\\[1ex]
$\psi(1,6,\bar{6},1)\sim 3(\bar{3},1,-4/3)\oplus 3(\bar{3},1,2/3)\oplus 
6(1,1,2)\oplus 9(1,1,0)\oplus 3(1,1,-2)$\\[1ex]
$\psi(6,\bar{6},1,1)\sim 9(1,2,1)\oplus 9(1,2,-1)$\\[1ex]
$\psi(1,1,6,\bar{6})\sim (8+1,1,0)\oplus 2(3,1,4/3)\oplus 2(\bar{3},1,-4/3)
\oplus (3,1,-2/3)\oplus (\bar{3},1,2/3)\oplus 5(1,1,0)\oplus 2(1,1,2)
\oplus 2(1,1,-2),$\\[1ex]
where the ordinary left-handed quarks correspond to 3(3,2,1/3) in 
$\psi(\bar{6},1,1,6)$, the ordinary right-handed quarks correspond to  
3($\bar{3},1,-4/3)\oplus 3(\bar{3},1,2/3)$ in $\psi(1,6,\bar{6},1)$,  
the known left-handed leptons are in three of the six (1,2,$-1$) of 
$\psi(\bar{6},1,1,6)$, and the known right-handed charged leptons are in 
three of the six (1,1,2) of $\psi(1,6,\bar{6},1)$. The 
exotic leptons in $\psi(\bar{6},1,1,6)$ belong to the vectorlike 
representation $3(1,2,-1)\oplus 3(1,2,1)$ (vectorlike with respect to 
the SM quantum numbers) and the exotic leptons in $\psi(1,6,\bar{6},1)$ 
belong to the vectorlike representation $3(1,1,2)\oplus 3(1,1,-2)\oplus 
9(1,1,0)$, where three lineal combinations of the nine states with 
quantum numbers (1,1,0) could be identified as the 
right-handed neutrinos.

$\psi(6,\bar{6},1,1)$ is formed by 36 exotic spin 1/2 Weyl fermions 
(we call them {\it nones} because they have zero lepton and baryon 
numbers), 9 with 
positive electric charges, 9 with negative (the charge conjugates to 
the positive ones), and 18 are neutrals; all together constitute a 
vectorlike representation with respect to the SM. 

Also all the particles in $\psi(1,1,6,\bar{6})$ form a vectorlike 
representation with respect to the SM, where 
$5(1,1,0)\oplus 2(1,1,2)\oplus 2(1,1,-2)$ stands for nine exotic 
fermions, five with zero electric charge ({\it nones}), two with  
electric charge +1 and the other two with electric charge $-1$ (spin 1/2 
dileptons); 
$2(3,1,4/3)\oplus 2(\bar{3},1,-4/3)$ refers to two exotic spin 1/2 
leptoquarks with 
electric charge 2/3; $(3,1,-2/3)\oplus (\bar{3},1,2/3)$ refers to one 
exotic spin 1/2 leptoquark with electric charge $-1/3$, and the 
nine states in (8+1,1,0)=(8,1,0)+(1,1,0) ($quaits$)+($quone$) are the 
so-called dichromatic fermion multiplets~\cite{gf} (also {\it nones}) 
which belong to the $(3,\bar{3})$ representation of the 
SU(3)$_{CR}\otimes$SU(3)$_{CL}$ subgroup of SU(6)$_{CR}\otimes$ SU(6)$_{CL}$.

Notice that contrary to the original Pati-Salam model, 
the $G$ symmetry and the representation content of 
$\psi(144)$ forbid mass terms for fermion fields 
at the unification scale, but according to 
the survival hypothesis\cite{sh} the vectorlike substructures pointed in this 
section (all the exotic particles in the model)
should get masses at scales above M$_Z$, the known weak interaction mass scale.

\section{\uno The Model}
The model under consideration contains 140 spin 1 gauge boson fields, 144 
spin 1/2 Weyl fermion fields, and a conveniently large number of spin 0 
scalar fields. We use for them the following notation:
\subsection{\uno The gauge bosons}
For the gauge boson fields we define:\\
a)-For the 70 gauge fields of SU(6)$_{CL}$ and SU(6)$_{CR}$ 
\begin{equation}
{\bf A}_{CL(CR)}=\frac{1}{\sqrt{2}}\left( 
\begin{array}{cccccc}
D_1  & G^1_2 & G^1_3 & \stackrel{\sim}{X_1} &  \stackrel{\sim}{Y_1} &
\stackrel{\sim}{Z_1}\\
G^2_1  & D_2 & G^2_3 & \stackrel{\sim}{X_2} &  \stackrel{\sim}{Y_2} &
\stackrel{\sim}{Z_2}\\
G^3_1  & G^3_2 & D_3 & \stackrel{\sim}{X_3} &  \stackrel{\sim}{Y_3} &
\stackrel{\sim}{Z_3}\\
X_1 & X_2 & X_3 & D_4 & P^-_1 & P^0 \\
Y_1 & Y_2 & Y_3 & P_1^+ & D_5 & P^+_2 \\
Z_1 & Z_2 & Z_3 & \stackrel{\sim}{P^0} & P^-_2 & D_6  
\end{array} \right)_{CL(CR)} \label{aclr}
\end{equation}
\noindent
where 
\begin{eqnarray*}
 {D_\delta}_{CL(CR)}&=&(G^\delta_\delta)_{CL(CR)} + 
\sqrt{\frac{1}{30}}B_{(B-L)_{L(R)}}+ 
\sqrt{\frac{2}{15}}B_{1YL(R)},\quad \delta=1,2,3;\\
{D_4}_{CL(CR)}&=&-\sqrt{\frac{3}{10}}B_{(B-L)_{L(R)}}-
\frac{1}{\sqrt{30}}B_{1YL(R)}-
\frac{1}{\sqrt{2}}B_{2YL(R)};\\
{D_5}_{CL(CR)}&=&\sqrt{\frac{3}{10}}B_{(B-L)_{L(R)}}-
\frac{4}{\sqrt{30}}B_{1YL(R)};\\
{D_6}_{CL(CR)}&=&-\sqrt{\frac{3}{10}}B_{(B-L)_{L(R)}}-
\frac{1}{\sqrt{30}}B_{1YL(R)}+
\frac{1}{\sqrt{2}}B_{2YL(R)};
\end{eqnarray*}
with $(G^\delta_\eta)_{CL(CR)}$, $\delta,\eta=1,2,3$ 
the gauge fields associated with SU(3) $_{CL(CR)}$  
($G^1_{1CL(CR)}$ $=B_{1gCL(CR)}/\sqrt{2}$ $+B_{2gCL(CR)}/\sqrt{6}$, 
$G^2_{2CL(CR)}=-B_{1gCL(CR)}/\sqrt{2}+B_{2gCL(CR)}/\sqrt{6}$, 
$G^3_{3CL(CR)}=-2B_{2gCL(CR)}/\sqrt{6}$ such that
$\sum_\delta (G^\delta_\delta)_{CL(CR)}=0$,  
and $B_{1gCL(CR)}$ and $B_{2gCL(CR)}$ are the gauge fields associated with the 
diagonal generators of  SU(3)$_{CL(CR)}$).
$B_{(B-L)_{L(R)}}$ is the gauge boson associated with the generator 
$Y_{(B-L)_{L(R)}}$, and $B_{1YL(R)}$ and $B_{2YL(R)}$ 
are two gauge bosons associated with the SU(6)$_{CL(CR)}$ diagonal 
generators $Y_{1L(R)}=Diag(2,2,2,-1,-4,$ $-1)/\sqrt{15}$ and 
$Y_{2L(R)}=Diag(0,0,0,-1,0,1)$ respectively.
$X_\delta, Y_\delta$ and $Z_\delta$ are spin 1 leptoquark gauge bosons 
with electric charges $-2/3,1/3$ and $-2/3$ respectively, with $\delta=1,2,3$ 
a color index. 
$P^{\pm}_\kappa, \kappa=1,2$; and $P^0$ are spin 1 dilepton gauge bosons 
with electric charges as indicated.\\
b)-For the 70 gauge fields of SU(6)$_L$ and SU(6)$_R$ 
\begin{equation}
{\bf A}_{L(R)}=\frac{1}{\sqrt{2}}\left( 
\begin{array}{cccccc}
A_1    & B^{\prime +}_1 & H^{\prime 0}_1 & B^+_2 &  H^0_2 & B^+_3 \\
B^{\prime -}_1  & A_2   & B^-_4 & H^{\prime 0}_3 &  B^-_5 & H^0_4 \\
\stackrel{\sim}{H}^{\prime 0}_1  & B^+_4 & A_3 & B^{\prime +}_6 
&  H^{\prime 0}_5 & B^+_7 \\
B^-_2  & \stackrel{\sim}{H}^{\prime 0}_3 & B^{\prime -}_6 & A_4 
& B^-_8  & H_6^{\prime 0} \\
\stackrel{\sim}{H}^0_2 & B^+_5 & \stackrel{\sim}{H}^{\prime 0}_5 & B^+_8 
& A_5 & B^{\prime +}_9 \\
B^-_3 & \stackrel{\sim}{H}^0_4 & B^-_7 & \stackrel{\sim}{H}_6^{\prime 0} 
& B^{\prime -}_9 & A_6  
\end{array} \right)_{L(R)} \label{3}
\end{equation}
\noindent
where the diagonal and the primed entries in Eq.(\ref{3}) are related to the 
physical fields as explained in the Appendix B.

\subsection{\uno The Fermionic content}
For the spin 1/2 Weyl fields we use the following definitions: 
\begin{equation}
\psi(\bar{6},1,1,6)_L=\left( 
\begin{array}{cccccc}
d_1  & d_2 & d_3 & e_{11}^-   & e_{12}^{0c} & e_{13}^- \\
u_1  & u_2 & u_3 & -n_{11}^0  & n_{12}^+    & -n_{13}^0 \\ 
s_1  & s_2 & s_3 & -e_{21}^-  & e_{22}^{0c} & e_{23}^- \\ 
c_1  & c_2 & c_3 & n_{21}^0   & n_{22}^+    & -n_{23}^0 \\ 
b_1  & b_2 & b_3 & -e_{31}^-  & e_{32}^{0c} & -e_{33}^- \\ 
t_1  & t_2 & t_3 & n_{31}^0   & n_{32}^+    & n_{33}^0  
\end{array} \right)_L \equiv\psi^\alpha_a 
\end{equation}
\noindent
where the rows(columns) represent color(flavor) degrees of freedom, 
$(u,d,c,s,b,t)$ are the quark fields with colors $\delta= 1,2,3$ as indicated, 
$(e_{ij},n_{ij}),\ i,j=1,2,3$ are lepton Weyl fields with electric charge 
as indicated, the minus signs are phases chosen for convenience,
 and the upper $c$ symbol stands for charge conjugation. 

\begin{equation}
\psi(1,6,\bar{6},1)=\left(
\begin{array}{cccccc}
d^c_1  & u^c_1 & s^c_1 & c^c_1 &  b^c_1 & t^c_1 \\
d^c_2  & u^c_2 & s^c_2 & c^c_2 &  b^c_2 & t^c_2 \\
d^c_3  & u^c_3 & s^c_3 & c^c_3 &  b^c_3 & t^c_3 \\
E^+_{11} & -N^{0c}_{11} & -E^+_{21} & N^{0c}_{21} & -E^+_{31}  & N^{0c}_{31} \\
E^0_{12} &  N^-_{12} &  E^0_{22} & N^-_{22} &  E^0_{32}  & N^-_{32} \\
E^+_{13} & -N^{0c}_{13} &  E^+_{23} & -N^{0c}_{23}& -E^+_{33}  & N^{0c}_{33} \\
\end{array} \right)_L \equiv\psi^A_\Delta 
\end{equation}
\noindent
where the rows (columns) now represent flavor (color) degrees of freedom. The
notation  we are using with the lepton fields in $\psi(1,6,\bar{6},1)$
unrelated in principle  to the lepton fields in $\psi(\bar{6},1,1,6)$ is
consistent with the SM quantum  numbers for
$\psi(\bar{6},1,1,6)\oplus\psi(1,6,\bar{6},1)$ presented in the Introduction. 
The known 
leptons $(\nu_e,e^-,\nu_\mu,\mu^-,\nu_\tau,\tau^-)$ and the known quarks  
are linear combinations of the leptons and quarks in 
$\psi(\bar{6},1,1,6)$ $\oplus\psi(1,6,\bar{6},1)$, up to mixing with exotics. 
Our notation is such that 
$a,b,..$; $A,B,...$; $\alpha,\beta,...$; $\Delta,\Omega,...$ stand for
SU(6)$_L$,  SU(6)$_R$, SU(6)$_{CL}$, and SU(6)$_{CR}$ tensor indices
respectively.

For the sake of completeness we also write:
\begin{equation}
\psi(1,1,6,\bar{6})\equiv\psi^\Delta_\alpha=\left( 
\begin{array}{cccccc}
g^1_1  & g^1_2 & g^1_3 & x_r &  y_r & z_r \\
g^2_1  & g^2_2 & g^2_3 & x_y &  y_y & z_y \\
g^3_1  & g^3_2 & g^3_3 & x_b &  y_b & z_b \\
\stackrel{\sim}{x_r} & \stackrel{\sim}{x_y} & \stackrel{\sim}{x_b} & 
l_1^0 & l^+_1 & l_2^0 \\
\stackrel{\sim}{y_r} & \stackrel{\sim}{y_y} & \stackrel{\sim}{y_b} & 
l_1^- & l_3^0 & l^-_2 \\
\stackrel{\sim}{z_r} & \stackrel{\sim}{z_y} & \stackrel{\sim}{z_b} & 
l^0_4 & l^+_2 & l_5^0 
\end{array} \right)_L 
\end{equation}
\noindent
where $g^i_j,\ i,j=1,2,3$ are the $(quaits)+(quone)$ spin 1/2 $nones$; 
$x,y$ and $z$ are the spin 1/2 leptoquarks with electric charges 2/3, 
$-1/3$ and 2/3 respectively, $l_j^{\pm}, j=1,2$ are spin 1/2 dilepton 
fields with electric charges as indicated, and $l_j^0,j=1,...5$ are five 
nones with zero electric charge.

\subsection{\uno The Scalar Content}
In order to spontaneously break the $G$ symmetry down to 
SU(3)$_C\otimes$U(1)$_{EM}$, and to implement at the same time the survival 
hypothesis and the horizontal survival hypothesis, we need to introduce 
the following rather complicated scalar sector:

First we introduce the scalar fields $\phi_1$ and $\phi_2$ with Vacuum 
Expectation Values (VeVs) such that 
$\langle\phi_1\rangle\sim\langle\phi_2\rangle\sim M$, where 

\[ \phi_j=\phi_j(900)=Z_4\phi_j(15,1,1,\overline{15})=
\phi^{[a,b]}_{j[\alpha,\delta]}+\phi^{[\alpha,\delta]}_{j[\Delta,\Omega]}+
\phi^{[\Delta,\Omega]}_{j[A,B]}+\phi^{[A,B]}_{j[a,b]} 
\] 

\noindent
$j=1,2$, and [.,.] stands for the commutator of the indices inside the 
brackets. The VeVs for $\phi_j,\ j=1,2$ are conveniently chosen in the 
following directions:\\[1ex]
$\langle\phi^{[a,b]}_{1[\alpha,\delta]}\rangle=\sqrt{3}M$ for 
$[a,b]=[4,1]=[2,3]=[5,6]; [\alpha,\delta]=[5,6]$\\[1ex]
$\langle\phi^{[\Delta,\Omega]}_{1[A,B]}\rangle=\sqrt{3}M$ for 
$[A,B]=[4,1]=[2,3]=[5,6]; [\Delta,\Omega]=[5,6]$\\ [1ex]
$\langle\phi^{[A,B]}_{1[a,b]}\rangle=M$ for $[a,b]=[A,B]=[4,1]=[2,3]=[6,5]$
\\[1ex]
$\langle\phi^{[\alpha,\delta]}_{j[\Delta,\Omega]}\rangle=0; j=1,2$\\[1ex]
$\langle\phi^{[a,b]}_{2[\alpha,\delta]}\rangle=\sqrt{3}M$ for 
$[a,b]=[1,2]=[6,3]=[4,5]; [\alpha,\delta]=[4,5]$\\[1ex]
$\langle\phi^{[\Delta,\Omega]}_{2[A,B]}\rangle=\sqrt{3}M$ for 
$[A,B]=[1,2]=[6,3]=[4,5]; [\Delta,\Omega]=[4,5]$\\ [1ex]
$\langle\phi^{[A,B]}_{2[a,b]}\rangle=M$ for $[a,b]=[A,B]=[2,1]=[6,3]=[4,5]$.
\\[1ex]
\indent It is easy to show\cite{abdel} that 
$\langle\phi_1\rangle+\langle\phi_2\rangle$ with the VeVs as indicated 
breaks 
\begin{center}
$G\longrightarrow$
SU(2)$_L\otimes$SU(2)$_R\otimes$SU(3)$_{CL}\otimes$SU(3)$_{CR}\otimes$
U(1)$_{(B-L)_L}\otimes$U(1)$_{(B-L)_R}$,
\end{center}

\noindent
the chiral extension of the left-right symmetric extension of the SM.

Next we introduce 

\[ \phi_3=\phi_3(5184)=Z_4\phi_3(1,1,(\overline{15}+\overline{21}),(15+21))=
\phi^{ab}_{3,\alpha\eta}+\phi^{\alpha\eta}_{3,\Delta\Omega}+
\phi^{\Delta\Omega}_{3,AB}+\phi^{AB}_{3,ab} \] 

\noindent
with the following VEVs:\\
$\langle\phi^{\alpha\eta}_{3,\Delta\Omega}\rangle=
M_C\delta^\alpha_\Omega\delta^\eta_\Delta; \hspace{.8cm} 
\alpha,\eta,\Delta,\Omega=1,..,6$,\\[1ex]
$\langle\phi^{[\Delta,\Omega]}_{3,[A,B]}\rangle=M_R$ for 
$[\Delta,\Omega]=\Delta\Omega-\Omega\Delta=[A,B]=[4,6]$,\\[1ex]
$\langle\phi^{ab}_{3,\alpha\eta}\rangle=\langle\phi^{AB}_{3,ab}\rangle=0.$

It then follows that 
\begin{center}
SU(6)$_{CR}\otimes$SU(6)$_{CL}
\stackrel{\langle\phi^{\alpha\eta}_{3,\Delta\Omega}\rangle}{\longrightarrow}$
SU(6)$_{(CL+CR)}\equiv$SU(6)$_C^V$,\\
\end{center}
\noindent
and that 
the main effect of $\langle\phi^{[\Delta,\Omega]}_{3,[A,B]}\rangle$ is to 
break SU(2)$_R\otimes$U(1)$_{(B-L)_R}$ in an appropriate way as we will 
shortly show.

Finally we introduce 

\[\phi_4=\phi_4(2592)=\phi_4(6,\bar{6},6,\bar{6})+
\phi_4(\bar{6},6,\bar{6},6)=\phi^{a\Omega}_{4,A\alpha}
+\phi_{4,a\Omega}^{A\alpha} \]

\noindent
with the following VEVs: 
$\langle\phi_{4,a\Omega}^{A\alpha}\rangle=0$, and 
$\langle\phi^{a\Omega}_{4,A\alpha}\rangle=M_Z$ for (a,A) = (6,6); 
$(\Omega,\alpha)$ = (1,1) = (2,2) = (3,3) = (4,4) = (5,5) = (6,6); and also 
for (a,A)=$(\Omega,\alpha)$=(5,5). As we will show in the next section 
the main effect of $\langle\phi_4\rangle$ is to break 
SU(2)$_L\otimes$U(1)$_Y$ down to U(1)$_{EM}$.

\section{\uno Tree level masses}
The scalar fields and their VEVs introduced in the previous section allow 
for the following tree level masses:
\subsection{\uno Masses for gauge bosons}
A tedious calculation\cite{abdel} in the sector of the covariant derivative in the 
Lagrangian shows the following results:\\
1. $\langle\phi_1\rangle+\langle\phi_2\rangle$ produces:
\begin{eqnarray}
{\cal L}(M)&=&g^2M^2\left\{18
\left[\sum_{\delta=1}^3\bigg(|X_{\delta CL}|^2 + 2|Y_{\delta CL}|^2 
+ |Z_{\delta CL}|^2\bigg) +2|P^0_{CL}|^2 \right.\right.\label{6} \\ \nonumber 
&+& |P_{1CL}|^2 + |P_{2CL}|^2 +\frac{5}{3}B^{2}_{1YL} 
+ B^{2}_{2YL}\left] + 24\left[\sum_{i=1}^8 c_i|B_{iL}|^2 + \sum_{i=1}^6 
c_i^\prime|H^0_{iL}|^2\right]\right.\\ \nonumber  
&+& 12\bigg(3A^2_{1HL}+A^{2}_{2HL} + A^2_{1AL} 
+3A^{2}_{2AL}\bigg) + (L\longrightarrow R)\Bigg\}, 
\end{eqnarray}
where $g$ is the gauge coupling constant for the simple group $G$, 
and the coefficients $c_i$ and $c_i^\prime$ are 
such that $c_1=c_2=c_3/2=c_4=c_5/2=c_6/3=c_7=c_8=1$ and 
$c_1^\prime/3=c_2^\prime/2=c_3^\prime=c_4^\prime/2=c_5^\prime/3=
c_6^\prime=1$.
(The relationship between the unprimed fields in Eq. (\ref{6}) and the primed 
ones in Eq. (\ref{3}) is presented in Appendix B.)

As it is clear from the former equation, 
$\langle\phi_1\rangle+\langle\phi_2\rangle$ breaks $G$ down 
to the chiral extension of the left-right symmetric extension of the SM.

\noindent
2. For $\langle\phi_3\rangle$ we split the analysis.\\
2a. $\langle\phi^{\alpha\eta}_{3,\Delta\Omega}\rangle$ produces:
\begin{eqnarray}
{\cal L}(M_C)&=&12g^2M_C^2
Tr\left[{\bf A}_{CL}^2-2{\bf A}_{CL}{\bf A}_{CR}+{\bf A}_{CR}^2\right]
 \label{7}\\ \nonumber 
&=&6g^2M^2_C\left[2\sum_{\delta=1}^3
\bigg(|X_{\delta CL}-X_{\delta CR}|^2 +|Y_{\delta CL}-Y_{\delta CR}|^2 
+|Z_{\delta CL}-Z_{\delta CR}|^2\bigg)\right. \\ \nonumber
&+&\sum_{i=1}^2 \bigg(2|P_{iCL}-P_{iCR}|^2 + \left(B_{igL}-B_{igR}\right)^2 + 
\left(B_{iYL}-B_{iYR}\right)^2\bigg)\\ \nonumber 
&+& \left(B_{(B-L)_L}-B_{(B-L)_R}\right)^2 + 2|P^0_{CL}-P^0_{CR}|^2 \\ \nonumber
&+&2\bigg(|G^1_{2CL}-G^1_{2CR}|^2+ |G^1_{3CL}-G^1_{3CR}|^2 
+ |G^2_{3CL}-G^2_{3CR}|^2\bigg)\Bigg].
\end{eqnarray}
As it is clear from the former expression,  
$\langle\phi^{\alpha\eta}_{3,\Delta\Omega}\rangle$ breaks 
SU(6)$_{CL}\otimes$SU(6)$_{CR}\longrightarrow$SU(6)$^V_C$ as mentioned 
before.

\noindent
2b. $\langle\phi^{[\Delta,\Omega]}_{3,[A,B]}\rangle$ for 
$[\Delta,\Omega]=[A,B]=[4,6]$ produces:
\begin{eqnarray}
{\cal L}(M_R)&=&2g^2M^2_R\left[\sum_{\delta=1}^3 
\bigg(|X_{\delta CR}|^2 + |Z_{\delta CR}|^2 \bigg) \right.
+ |P_{1 CR}^+|^2 + |P_{2 CR}^+|^2  \nonumber \\
&+& |B_{2R}|^2 + |B_{3R}|^2 + |B^\prime_{6R}|^2 + |B_{7R}|^2  
+ |B_{8R}|^2 + |B^\prime_{9R}|^2 \nonumber \\
&+& |H^\prime_{3R}|^2 + |H_{4R}|^2 + 
\left.\left.{4\over 3}\right({\bf B}^0_{CR}-{\bf W}^0_R\right)^2\Bigg]   
\end{eqnarray}
\noindent
where 
\[ {\bf B}^0_{CR}=\left(3B_{(B-L)_R}+B_{1YR}\right)/\sqrt{10} \] 
\noindent
and
\[{\bf W}^0_R=\left(\sqrt{8}W^0_R +\sqrt{3}A_{1HR} + A_{2HR} - \sqrt{3}A_{1AR} 
- A_{2AR}\right)/\sqrt{16}.\]
\noindent
The mixing between SU(6)$_{CR}$ and SU(6)$_R$ is given by 
${\bf B}^0_{CR}\cdot{\bf W}^0_R$. The analysis shows also that 
$\langle\phi^{[\Delta,\Omega]}_{3,[A,B]}\rangle$, with the VEVs as stated 
breaks SU(6)$_{CR}\otimes$SU(6)$_R$ down to \\
SU(4)$^{\prime\prime}_{CR}\otimes$SU(2)$^{\prime\prime}_{CR}
\otimes$SU(4)$^{\prime\prime}_{R}\otimes$SU(2)$^{\prime\prime}_{R}
\otimes$U(1)$_{mix}$, where U(1)$_{mix}$ is associated with the unbroken 
gauge boson  $({\bf B}^0_{CR}+{\bf W}^0_R)/\sqrt{2}$. 

It is also a matter of a careful analysis to realize that 
$\langle\phi_1\rangle +\langle\phi_2\rangle + \langle\phi_3\rangle$  breaks 
$G$ down to SU(3)$_C\otimes$SU(2)$_L\otimes$U(1)$_Y$, the gauge group of the SM.

3. $\langle\phi_4\rangle$ with the VEVs as indicated produces:
\begin{eqnarray}
{\cal L}(M_Z)&=&g^2M^2_Z
\Big\{Tr\left[{\bf A}^2_{CL} -2{\bf A}_{CL}{\bf A}_{CR}+ {\bf A}^2_{CR} 
+{\bf A}^2_L I_5^2 - 2{\bf A}_L I_5{\bf A}_R I_5 
+ {\bf A}^2_R I_5^2 \right. \nonumber  \\
&+& \left. 6\left({\bf A}^2_L I_6^2 - 2{\bf A}_L I_6{\bf A}_R I_6 + 
{\bf A}^2_R I_6^2\right)
+{\bf A}^2_{CL} I_5^2 - 2{\bf A}_{CL}I_5{\bf A}_{CR}I_5 
+ {\bf A}^2_{CR} I_5^2\right]  \nonumber  \\
&-&\left. 2\left({\bf A}_L-{\bf A}_R\left)^5_5
\right({\bf A}_{CL}-{\bf A}_{CR}\right)^5_5\right\} \nonumber \\
&=&\frac{M_Z^2}{12M_C^2}{\cal L}(M_C)
+{\cal L}^{LCR}_{\langle\phi_4\rangle}, 
\end{eqnarray}
\noindent
where ${\cal L}(M_C)$ is 
given by Eq. (\ref{7}), and $I_5=\delta_{55}$ and $I_6=\delta_{66}$ are 
$6\times 6$ matrices with only one entry different from zero, which produce:
\begin{eqnarray}
{\cal L}^{LCR}_{\langle\phi_4\rangle}&=&{g^2M_Z^2\over 2}
\Bigg[6|B_{3L}|^2+|B_{5L}|^2+6|B_{7L}|^2+|B_{8L}|^2+7|B^\prime_{9L}|^2
+|H^0_{2L}|^2 +6|H^0_{4L}|^2 \nonumber \\
&+&|H^{\prime 0}_{5L}|^2 +6|H^{\prime 0}_{6L}|^2 
+\left(\sum_{\delta = 1}^3|Y_{\delta CL}|^2\right)+|P_{1L}|^2+|P_{2L}|^2 
+(L\longrightarrow R) \nonumber \\
&+& 6\big(A_{6L}-A_{6R}\big)^2 + \big(A_{5L}-D_{5L}-A_{5R}+D_{5R}\big)^2\Bigg]
\end{eqnarray}

Combining the former equations we see that the only gauge bosons that remain
massless are:\\
1. The eight fields
$G^{V,\delta}_\eta=(G^\delta_{\eta,CL}+G^\delta_{\eta,CR})/\sqrt{2},
\delta,\eta=1,2,3, (\sum_\delta G^{V,\delta}_\delta=0$), associated with the
gauge bosons for SU(3)$_C$.\\
2. $A=\frac{3}{\sqrt{28}}[W^0_L+W^0_R-\frac{\sqrt{5}}{3}(B_{(B-L)_L}+B_{(B-L)_R})]$
which is the photon field. Then using the
identity $A=sin\theta_W W^0_L + cos\theta_W B_Y$, where $\theta_W$ is the weak
mixing angle, we get $sin\theta_W=3/\sqrt{28}$ at the G scale, and 
$B_Y=[3W^0_R-\sqrt{5}(B_{(B-L)_L}+B_{(B-L)_R})]/\sqrt{19}$ as the boson
associated with U(1)$_Y$.


\subsection{\uno Masses for fermion fields}
With the scalar fields of the model $\phi_i, i=1,...,4$ we can construct the 
following Yukawa terms:

\[Z_4\psi(\bar{6},1,1,6)\psi(\bar{6},1,1,6)[\sum_{i=1}^2y_i
\phi_i(15,1,1,\overline{15}) + y_3\phi_3((15+21),1,1,
(\overline{15}+\overline{21}))]\] 
\[+ y_4Z_4\psi(\bar{6},1,1,6)\psi(1,6,\bar{6},1)\phi_4(6,\bar{6},6,\bar{6}) 
+ h.c. \]

\noindent
where $y_i,i=1,...4$ are Yukawa coupling constants of order one. When the 
Higgs fields $\phi_i,i=1-4$ develop the VEVs as indicated in Section 2.3 
they produce the following masses for the fermion fields:

\subsubsection{\uno Masses from $\langle\phi_1\rangle+\langle\phi_2\rangle$}
$Z_4\psi(\bar{6},1,1,6)\psi(\bar{6},1,1,6)\sum_{i=1}^2y_i
\langle\phi_i(15,1,1,\overline{15})\rangle$ produces\\
1. Masses of order M for all the exotic {\it nones} in 
$\psi(6,\bar{6},1,1)$.\\ 
2. The following Dirac masses: 
\begin{eqnarray}
{\cal L}_M^{\langle\phi_1\rangle+\langle\phi_2\rangle}
&=& n_{12}^+(Y_1e_{23}^-+Y_2e_{11}^-)+
n_{22}^+(Y_1e_{13}^-+Y_2e_{31}^-)+  
n_{32}^+(Y_1e_{33}^-+Y_2e_{21}^-)  \nonumber \\
&+& N_{12}^-(Y_1E_{23}^++Y_2E_{11}^+)+
N_{22}^-(Y_1E_{13}^++Y_2E_{31}^+)+  
N_{32}^-(Y_1E_{33}^++Y_2E_{21}^+)  \nonumber \\
&+& E_{12}^{0}(Y_1N_{23}^{0c}+Y_2N_{11}^{0c})+  
E_{22}^{0}(Y_1N_{13}^{0c}+Y_2N_{31}^{0c}) + 
E_{32}^{0}(Y_1N_{33}^{0c}+Y_2N_{21}^{0c}) \nonumber \\
&+& e_{12}^{0c}(Y_1n_{23}^{0}+Y_2n_{11}^{0})+  
e_{22}^{0c}(Y_1n_{13}^{0}+Y_2n_{31}^{0}) 
+ e_{32}^{0c}(Y_1n_{33}^{0}+Y_2n_{21}^{0})\nonumber\\ &+& h.c.,
\label{phi12}
\end{eqnarray}

\noindent
where $Y_i=\sqrt{3}M y_i$, $i=1,2$.
   Equation (\ref{phi12}) allows us to 
identify  $\kappa(Y_2e_{23}-Y_1e_{11})$, $\kappa (Y_2e_{13}-Y_1e_{31})$ and  
$\kappa(Y_2e^-_{33}-Y_1e^-_{21})$ with $\kappa=(Y_1^2+Y_2^2)^{-1/2}$ as a 
basis for the known charged left-handed leptons $(e^-,\mu^-,\tau^-)_L$; 
$\kappa(Y_2E^+_{23}-Y_1E^-_{11})$, $\kappa (Y_2E_{13}^+-Y_1E_{31}^+)$ and   
$\kappa(Y_2E^+_{33}-Y_1E^-_{21})$ as a 
basis for the known charged right-handed leptons $(e^-,\mu^-,\tau^-)_R$;  
$\kappa(Y_2n_{23}^{0}-Y_1n_{11}^{0})$, 
$\kappa(Y_2n_{31}^{0}-Y_1n_{13}^{0})$ and 
$\kappa(Y_2n_{33}^{0}-Y_1n_{21}^{0})$ as a basis for 
$(\nu_e,\nu_\mu,\nu_\tau)_L$, and 
$\kappa(Y_2N_{23}^{0c}-Y_1N_{11}^{0c})$, 
$\kappa(Y_2N_{31}^{0c}-Y_1N_{13}^{0c})$ and 
$\kappa(Y_2N_{33}^{0c}-Y_1N_{21}^{0c})$ as a basis for 
$(\nu^c_e,\nu^c_\mu,\nu^c_\tau)_L$. 

As can be seen from the former expression, all the vector-like particles with
respect to the chiral extension of the left-right symmetric extension of the SM
acquire masses of order $M$, as it should be according to the survival
hypothesis\cite{sh} (see Appendix A).

\subsubsection{\uno Masses from $\langle\phi_3\rangle$}
$Z_4\psi(1,1,6,\bar{6})\psi(1,1,6,\bar{6})
\langle\phi_3(1,1,(\overline{15}+\overline{21}),(15+21))\rangle$ 
producces the following masses:\\
1. Dirac masses for all the exotic fields in $\psi(1,1,6,\bar{6})$ of 
order $M_C$, via the Yukawa term $y_3\psi^\Delta_\alpha\psi^\Omega_\eta
\langle\phi^{\alpha\eta}_{3,\Delta\Omega}\rangle$\\
2. The following Majorana masses:
\begin{eqnarray}
{\cal L}_{M_R}^{\langle\phi_3\rangle}&=&y_3\psi_\Delta^A\psi_\Omega^B
\langle\phi_{3,[A,B]}^{[\Delta,\Omega]}\rangle  \nonumber \\
&=&y_3M_R(N^{0c}_{21L}N^{0c}_{33L}+N^{0c}_{23L}N^{0c}_{31L}
+N^{0c}_{33L}N^{0c}_{21L}+N^{0c}_{31L}N^{0c}_{23L}),
\end{eqnarray}

\subsubsection{\uno Masses from $\langle\phi_4\rangle$}
$\phi_4$, with the VEVs as stated in the previous section, produces the 
following mass terms:
\begin{eqnarray}
\label{mtop}
{\cal L}_{M_Z}^{\langle\phi_4\rangle}&=& 
y_4\left(\psi^\alpha_a\psi^A_\Delta + 
\psi^A_\Delta\psi^\alpha_a\right)\langle\phi^{a\Delta}_{4,\alpha A}\rangle 
\\ \nonumber
&=&y_4M_Z\left[\sum_{\delta=1}^3t^c_{\delta L}t_{\delta L} 
+ N^{0c}_{31L}n^0_{31L} 
+N^{-}_{32L}n^+_{32L}+N^{0c}_{33L}n^0_{33L}+E^{0}_{32L}e^{0c}_{32L}
+h.c.
\right ],
\end{eqnarray}

\noindent
from where we can immediately see that the top quark (but not the bottom quark)
gets  a tree level mass $m_t=y_4M_Z$. The algebra also shows that
Eq.[\ref{mtop}] contains  a small mass term for one of the neutrino
fields\cite{pzab}. This is the way how we  achieve the modified  horizontal
survival hypothesis in the context of the model presented here.

\section{\uno Mass scales}
\subsection{\uno The electroweak mixing angle}
There are several ways to calculate the electroweak mixing angle at the
unification  scale ($M_G$) for a grand unified theory. For a simple gauge group
the relationship\cite{wein}
\[sin^2\theta_W(M_G)=tr(T^2_{ZL})/tr(Q^2),\]
\noindent
may be used, where the traces can be evaluated using any faithful 
representation
(reducible  or irreducible) of the simple group.

Now, [SU(6)]$^4$ is not simple, but [SU(6)]$^4\times Z_4$ is. Therefore 
we can calculate the traces for $\psi(144)$ and plug them in the 
former 
expression. Note that all the four sectors of $\psi(144)$ must be used in the 
computation of the traces due to the fact that a single sector is not a
faithful representation of G because it is not $Z_4$ invariant. 
After the algebra is done we get $sin^2\theta_W(M_G)=9/28$ in agreement with
the previous calculation, and the same value 
obtained for the three family extension of the Pati-Salam model with mirror 
fermions\cite{elias}.

Now, if we define $g_1,g_2$, and $g_3$ as the gauge coupling constants for
U(1)$_Y$, SU(2)$_L$, and SU(3)$_C$ respectively, the the embedding of the SM
model gauge group for three families in $[G,\psi(144)]$, and the former value
for $sin^2\theta_W$ imply that at the G scale the following relationships
holds\cite{pz,elias}: $g_3=g/\sqrt{2}, g_2=g/\sqrt{3}$, and $g_1=\sqrt{3/19}g$.
At scales well below the G scale the former relations are not longer valid
because the embedding symmetry G is not manifest, then the effective coupling
constants must be evaluated using the renormalization group equations.

\subsection{\uno The renormalization group equations}
Next we introduce the renormalization group equations and use standard
decoupling theorem arguments\cite{ac} in order to calculate the 
mass scales.

For generality, let us analyze the two mass scale symmetry breaking pattern 
\[G\stackrel{M_R=M_C}{\longrightarrow}G_I\stackrel{M}{\longrightarrow} 
SU(3)_C\otimes SU(2)_L\otimes U(1)_Y
\stackrel{M_Z}{\longrightarrow}SU(3)_C\otimes U(1)_{EM}\]
with $M_R>>M>>M_Z$, and
$G_I$ = SU(6)$_L\,\otimes$ SU(4)$^V_C\,\otimes$ U(1)$_Y\,\otimes...$, where
SU(3)$_C\,\subset$ SU(4)$^V_C$ and  SU(2)$_L\,\subset$ SU(6)$_L$. For this 
two-stage
gauge hierarchy the runing coupling constants of the SM satisfy the one loop
renormalization group equations\cite{saly}
 \begin{equation}
\alpha^{-1}_i(M_Z)=f_i\alpha^{-1}-b^{M_R}_i{\rm ln}\left({M_R\over M}\right) 
-b_i^M{\rm ln}\left({M\over M_Z}\right),  
\label{eq1}
\end{equation}
where $\alpha_i=g_i^2/4\pi$, $i=1,2,3, \alpha=g^2/4\pi$, and $f_i$ are
embedding constants given by $f_1=19/3, f_2=3$ and $f_3=2$. The beta functions
are:
 \begin{equation}
b_i=\{\frac{11}{3}C_i(vectors)
-\frac{2}{3}C_i(Weyl-fermions)-\frac{1}{6}C_i(scalars)\}/4\pi,
\label{eq2}
\end{equation}
where $C_i(...)$ is the index of the representation to which the
(...) particles are assigned, and the $C_i(Weyl-fermions)$ and $C_i(scalars)$
indexs must be properly normalized with the embedding factor $f_i$.

Now, using the relationships $e^{-2}=g_1^{-2}+g_2^{-2}$ and
tan$\theta_W=g_1/g_2$, valid at all energy scales, we get from 
Eqs(\ref{eq1}): 
 \begin{equation}
\frac{3}{28}\alpha^{-1}_{EM}(M_Z)=\frac{\alpha^{-1}_3(M_Z)}{2}+
(\frac{b^{M_R}_3}{2}-\frac{3b^{M_R}_2}{28}-
\frac{3b^{M_R}_1}{28}){\rm ln}\left({M_R\over M}\right) + 
(\frac{b^{M}_3}{2}-\frac{3b^{M}_2}{28}-
\frac{3b^{M}_1}{28}){\rm ln}\left({M\over M_Z}\right),  
\label{eq5}
 \end{equation}
and
 \begin{equation}
{\rm sin}^2\theta_W(M_Z)=3\alpha_{EM}(M_Z)
\{\frac{\alpha_3^{-1}(M_Z)}{2}+
(\frac{b^{M_R}_3}{2}-\frac{b^{M_R}_2}{3})
{\rm ln}\left({M_R\over M}\right) +
(\frac{b^{M}_3}{2}-\frac{b^{M}_2}{3})
{\rm ln}\left({M\over M_Z}\right)\}.
\label{eq6}
\end{equation}

\noindent
As it is well known, the Higgs fields play an important role in the beta
functions\cite{aguila} and can drastically change the solutions to the
renormalization group equations. So, we are going to solve those equations
under the assumption that the extended survival hypothesis holds\cite{aguila}.
Using this hypothesis,  decoupling the vector-like representations in 
$\psi(144)$  
according to the Appel\-quist--Carazzone theorem\cite{ac}, and using the 
experimental values\cite{data} $sin^2\theta_W(M_Z)=0.2319, 
\alpha_3(M_Z)=0.117$ and $\alpha_{EM}^{-1}(M_Z)=127.6$ we get the solutions
$M=5.0 \times 10^5M_Z$ and $M_R=5.5 M$. When the threshold effects and the
experimental errors are taken into account, the solution is compatible with
the amazing result $M_R=M_C\sim M\sim 10^8>>M_Z\sim 10_2$ GeVs, which
implies that only one stage symmetry breaking pattern is required, and there is
only one mass scale between the G and the electroweak scales.

So, our model is compatible with the symmetry breaking pattern: 
\[G\stackrel{M}{\longrightarrow}SU(3)_C\otimes SU(2)_L\otimes U(1)_Y 
\stackrel{M_Z}{\longrightarrow}SU(3)_C\otimes U(1)_{EM},\]
\noindent
where $M\sim 10^8$ GeVs, and $M_Z\sim 10^2$GeVs is the electroweak mass
scale. Notice also that the lower value of the G scale softens the gauge
hierarchy problem.

\section{\uno Interacting Lagrangian}
Using the covariant derivative for $G$ we can write the following 
interacting terms:
 \begin{eqnarray}
{\cal L}^{int}&=&g[\psi(\bar{6},1,1,6){\bf A}_{CL}\psi(\bar{6},1,1,6) 
-\psi(\bar{6},1,1,6){\bf A}_{L}\psi(\bar{6},1,1,6) \nonumber \label{lint} \\
&+&\psi(1,6,\bar{6},1){\bf A}_{R}\psi(1,6,\bar{6},1) 
-\psi(1,6,\bar{6},1){\bf A}_{CR}\psi(1,6,\bar{6},1)  \nonumber  \\
&+&\psi(1,1,6,\bar{6}){\bf A}_{CR}\psi(1,1,6,\bar{6}) 
-\psi(1,1,6,\bar{6}){\bf A}_{CL}\psi(1,1,6,\bar{6})  \nonumber  \\
&+&\psi(6,\bar{6},1,1){\bf A}_{L}\psi(6,\bar{6},1,1) 
-\psi(6,\bar{6},1,1){\bf A}_{R}\psi(6,\bar{6},1,1)] \nonumber \\
&\equiv &{\cal L}_{CL}+{\cal L}_{R}-{\cal L}_{L}-{\cal L}_{CR}.
 \end{eqnarray}
As far as the ordinary particles are concerned, each term in 
${\cal L}^{int}$ may be written as 
\[ {\cal L}_i={\cal L}_i^{qq} + {\cal L}_i^{ql} + {\cal L}_i^{ll} \]
for $i=CL,R,L,CL$, where $qq,ql$, and $ll$ stand for quark-quark, quark-lepton 
and lepton-lepton interactions respectively. Also for our concern here, only 
the terms in Eq. (\ref{lint}) with known fields must be evaluated explicitly.

After the algebra is done we get the following expressions:
\begin{equation}
{\cal L}_{CL}^{qq}=\frac{g}{\sqrt{2}}
\left\{\sum_{q=u,d,c,s,t,b}\left[\sum_{\delta\neq \eta=1}^3\bar{q}_{\delta L}
(G^\delta_\eta)^\mu_{CL}\gamma_{\mu} q_{\eta L}+
\sum_{\delta=1}^3\bar{q}_{\delta L}D^\mu_{\delta CL}
\gamma_\mu q_{\delta L}\right]\right\} ,
\label{liqq} \end{equation}
\begin{eqnarray}
{\cal L}_{CL}^{ql}&=&\frac{g}{\sqrt{2}}
\sum_{\delta=1}^3
\bigg[X^\mu_{\delta CL}\big(\bar{\bf n}_{1L}^0\!\cdot\!\gamma_\mu U_{\delta L}+
\bar{\bf e}_{1L}^-\!\cdot\!\gamma_\mu D_{\delta L}\big) + 
Y^\mu_{\delta CL}\big(\bar{\bf n}_{2L}^+\!\cdot\!\gamma_\mu U_{\delta L} + 
\bar{\bf e}_{2L}^{0c}\!\cdot\!\gamma_\mu D_{\delta L}\big)  \nonumber    \\ 
&+& Z^\mu_{\delta CL}\big(\bar{\bf n}_{3L}^0\!\cdot\!\gamma_\mu U_{\delta L} + 
\bar{\bf e}_{3L}^-\!\cdot\!\gamma_\mu D_{\delta L}\big) 
+ h.c.\bigg] ,   \label{liql} 
\end{eqnarray}
\begin{eqnarray}
{\cal L}_{CL}^{ll}&=&\frac{g}{\sqrt{2}}
\Big[D_{4CL}^\mu\big(\bar{\bf n}_{1L}^0\!\cdot\!\gamma_\mu {\bf n}_{1L}^0 + 
\bar{\bf e}_{1L}^-\!\cdot\!\gamma_\mu {\bf e}_{1L}^-\big) 
+ D_{5CL}^\mu\big(\bar{\bf n}_{2L}^+\!\cdot\!\gamma_\mu {\bf n}_{2L}^+  + 
\bar{\bf e}_{2L}^{0c}\!\cdot\!\gamma_\mu {\bf e}_{2L}^{0c}\big) 
\nonumber \label {lill}  \\
&+& D_{6CL}^\mu\big(\bar{\bf n}_{3L}^0\!\cdot\!\gamma_\mu {\bf n}_{3L}^0 
+ \bar{\bf e}_{3L}^-\!\cdot\!\gamma_\mu {\bf e}_{3L}^-\big)   
+ P_{CL}^{0,\mu}\big(\bar{\bf n}_{1L}^0\!\cdot\!\gamma_\mu {\bf n}_{3L}^0 
+ \bar{\bf e}_{1L}^-\!\cdot\!\gamma_\mu {\bf e}_{3L}^-\big) \nonumber   \\
&+& P_{1CL}^{+,\mu}\big(\bar{\bf n}_{2L}^+\!\cdot\!\gamma_\mu {\bf n}_{1L}^0 + 
\bar{\bf e}_{2L}^{0c}\!\cdot\!\gamma_\mu {\bf e}_{1L}^-\big) + 
P_{2CL}^{+,\mu}\big(\bar{\bf n}_{2L}^+\!\cdot\!\gamma_\mu {\bf n}_{3L}^0 + 
\bar{\bf e}_{2L}^{0c}\!\cdot\!\gamma_\mu {\bf e}_{3L}^-\big)  \nonumber \\ 
&+& h.c. \bigg] ,   
\end{eqnarray}
\noindent
where we have defined the following three component vectors: 
$U_\delta=(u_\delta,c_\delta,t_\delta)$, 
$D_\delta=(d_\delta,s_\delta,b_\delta)$;
${\bf n}_1^0=(-n_{11}^0,n_{21}^0,n_{31}^0)$,  
${\bf n}_2^+=(n_{12}^+,n_{22}^+,n_{32}^+)$, 
${\bf n}_3^0=(-n_{13}^0,-n_{23}^0,n_{33}^0)$;  
${\bf e}_1^-=(e_{11}^-,-e_{21}^-,-e_{31}^-)$, 
${\bf e}_2^{0c}=(e_{12}^{0c},e_{22}^{0c},e_{32}^{0c})$ and
${\bf e}_3^-=(e_{13}^-,e_{23}^-,-e_{33}^-)$.

${\cal L}^{qq}_{CR},  {\cal L}^{ql}_{CR}$ and ${\cal L}^{ll}_{CR}$ are 
expressions similar to the ones pressented in Eqs. (\ref{liqq}), (\ref{liql})
and (\ref{lill}) with the following changes: replacement of $CL\rightarrow CR$
in all the gauge  fields, changing of the quark fields $U_L$ and $D_L$ by
their  corresponding charge conjugated fields $U^c_L$ and $D^c_L$, and changing
of the lepton vectors ${\bf n}_1^0$, ${\bf n}_2^+$, ${\bf n}_3^0$, 
${\bf e}_1^-$, ${\bf e}_2^{0c}$ and ${\bf e}_3^-$  by 
${\bf N}_1^{0c}=(-N_{11}^{0c},N_{21}^{0c},N_{31}^{0c})$,  
${\bf N}_2^-=(N_{12}^-,N_{22}^-,N_{32}^+)$, 
${\bf N}_3^{0c}=(-N_{13}^{0c},-N_{23}^{0c},N_{33}^{0c})$;  
${\bf E}_1^+=(E_{11}^+,-E_{21}^+,-E_{31}^+)$, 
${\bf E}_2^0=(E_{12}^{0},$ $E_{22}^{0},$ $E_{32}^{0})$ and
${\bf E}_3^+=(E_{13}^+,E_{23}^+,-E_{33}^+)$, respectively.
 Then the right-handed 
fields will show up in the final expressions by using the identity 
$\bar{\chi}^c_L\gamma^\mu\xi^c_L=-\bar{\xi}_R\gamma^\mu\chi_R$. Then 
$-{\cal L}_{CR}$ will be just ${\cal L}_{CL}$ with the substitutions
$L\rightarrow R$  and $\{{\bf n},{\bf e}\}\rightarrow\{{\bf N},{\bf E}\}$
everywhere.

Next for ${\cal L}_L$ and ${\cal L}_R$ we have the result that 
${\cal L}^{ql}_{L(R)}=0$. Then ${\cal L}^{qq}_{L(R)}$ and 
${\cal L}^{ll}_{L(R)}$ can be conveniently written as:
\[ {\cal L}^{qq(ll)}_{L(R)} ={\cal L}^{q(l)W}_{L(R)} +
{\cal L}^{q(l)H}_{L(R)}+{\cal L}^{q(l)A}_{L(R)}+{\cal L}^{q(l)B}_{L(R)}.\] 
\noindent
After the algebra is done we get the following expressions
\begin{equation}
{\cal L}_{L}^{qW}=\frac{g}{\sqrt{6}}\sum_{\delta =1}^3\left[
\frac{W^0_{\mu L}}{\sqrt{2}}
\Big(\bar{D}_{\delta L}\!\cdot\!\gamma^{\mu} D_{\delta L} - 
\bar{U}_{\delta L}\!\cdot\!\gamma^\mu U_{\delta L}\Big) + 
\Big(W^+_{\mu L}\bar{U}_{\delta L}\!\cdot\!\gamma^\mu D_{\delta L} + h.c.\Big) 
\right],
\label{liqw} \end{equation}

\begin{eqnarray}
{\cal L}_{L}^{qH}&=&\frac{g}{\sqrt{2}}\sum_{\delta=1}^3
\bigg[\frac{H^{0\mu}_{1L}}{\sqrt{2}}
\big(\bar{c}_{\delta L}\gamma_\mu u_{\delta L} + 
\bar{s}_{\delta L}\gamma_\mu d_{\delta L}\big) + \label{liqh}
H^{0\mu}_{2L}\ \bar{b}_{\delta L}\gamma_\mu d_{\delta L}  \nonumber \\
&+&\frac{H^{0\mu}_{3L}}{\sqrt{2}}
\big(\bar{c}_{\delta L}\gamma_\mu u_{\delta L} - 
\bar{s}_{\delta L}\gamma_\mu d_{\delta L}\big) + 
H^{0\mu}_{4L}\ \bar{t}_{\delta L}\gamma_\mu u_{\delta L} \nonumber \\
&+&\frac{H^{0\mu}_{5L}}{\sqrt{2}}
\big(\bar{b}_{\delta L}\gamma_\mu s_{\delta L} + 
\bar{t}_{\delta L}\gamma_\mu c_{\delta L}\big) + 
\frac{H^{0\mu}_{6L}}{\sqrt{2}}
\left(\bar{t}_{\delta L}\gamma_\mu c_{\delta L} - 
\bar{b}_{\delta L}\gamma_\mu s_{\delta L}\right) + h.c.\bigg],
\end{eqnarray}
\begin{eqnarray}
{\cal L}_{L}^{qA}&=&\frac{g}{2\sqrt{2}}\sum_{\delta=1}^3
\bigg[A^\mu_{1HL}\big(\bar{d}_{\delta L}\gamma_\mu d_{\delta L} + 
\bar{u}_{\delta L}\gamma_\mu u_{\delta L} - 
\bar{b}_{\delta L}\gamma_\mu b_{\delta L} -  \label{liqa}
\bar{t}_{\delta L}\gamma_\mu t_{\delta L}\big)  \nonumber \\
&+&\frac{A^\mu_{2HL}}{\sqrt{3}}
\big(\bar{d}_{\delta L}\gamma_\mu d_{\delta L} + 
\bar{u}_{\delta L}\gamma_\mu u_{\delta L} -  
2\bar{s}_{\delta L}\gamma_\mu s_{\delta L} - 
2\bar{c}_{\delta L}\gamma_\mu c_{\delta L} + 
\bar{b}_{\delta L}\gamma_\mu b_{\delta L} + 
\bar{t}_{\delta L}\gamma_\mu t_{\delta L}\big) \nonumber \\
&+&\frac{A^\mu_{2AL}}{\sqrt{3}}
\big(\bar{d}_{\delta L}\gamma_\mu d_{\delta L} - 
\bar{u}_{\delta L}\gamma_\mu u_{\delta L} -  
2\bar{s}_{\delta L}\gamma_\mu s_{\delta L} + 
2\bar{c}_{\delta L}\gamma_\mu c_{\delta L} + 
\bar{b}_{\delta L}\gamma_\mu b_{\delta L} - 
\bar{t}_{\delta L}\gamma_\mu t_{\delta L}\big)  \nonumber \\
&+& A^\mu_{1AL}\big(\bar{d}_{\delta L}\gamma_\mu d_{\delta L} - 
\bar{u}_{\delta L}\gamma_\mu u_{\delta L} - 
\bar{b}_{\delta L}\gamma_\mu b_{\delta L} + 
\bar{t}_{\delta L}\gamma_\mu t_{\delta L}\big) \bigg],
\end{eqnarray} 

\begin{eqnarray}
{\cal L}_{L}^{qB}&=&\frac{g}{\sqrt{2}}\sum_{\delta=1}^3
\bigg[\frac{B^{-\mu}_{1L}}{\sqrt{2}}
\big(\bar{d}_{\delta L}\gamma_\mu u_{\delta L} - 
\bar{b}_{\delta L}\gamma_\mu t_{\delta L}\big)  \label{liqb}
+B^{-\mu}_{2L}\ \bar{d}_{\delta L}\gamma_\mu c_{\delta L} 
+B^{-\mu}_{3L}\ \bar{d}_{\delta L}\gamma_\mu t_{\delta L}  \nonumber \\
&+& B^{-\mu}_{4L}\ \bar{s}_{\delta L}\gamma_\mu u_{\delta L}
+B^{-\mu}_{5L}\ \bar{b}_{\delta L}\gamma_\mu u_{\delta L} + 
\frac{B^{-\mu}_{6L}}{\sqrt{6}}
\big(\bar{b}_{\delta L}\gamma_\mu t_{\delta L} - 
2\bar{s}_{\delta L}\gamma_\mu c_{\delta L} + 
\bar{d}_{\delta L}\gamma_\mu u_{\delta L}\big) \nonumber \\[1ex]
&+&B^{-\mu}_{7L}\ \bar{s}_{\delta L}\gamma_\mu t_{\delta L}
+B^{-\mu}_{8L}\ \bar{b}_{\delta L}\gamma_\mu c_{\delta L} + h.c.\bigg].
\end{eqnarray}
\noindent
Again, $-{\cal L}^{qq}_R$ is just ${\cal L}^{qq}_L$ with the substitution 
$L\rightarrow R$ everywhere. 

The expressions for ${\cal L}^{ll}_{L(R)}$ are  very similar to
${\cal L}^{qq}_{L(R)}$. In fact ${\cal L}^{ll}_L$ is just ${\cal L}^{qq}_L$
with the substitutions
$D_\delta\rightarrow {\bf e}_\delta$ and $U_\delta\rightarrow {\bf n}_\delta$,
in the expression for ${\cal L}_L^{qW}$, and 
$u_\delta\rightarrow {\bf{\eta}}_1=(-n^0_{11},n^+_{12},-n^0_{13})$, 
$c_\delta\rightarrow {\bf{\eta}}_2= (n^0_{21},n^+_{22},-n^0_{23})$, 
$t_\delta\rightarrow {\bf{\eta}}_3= (n^0_{31},n^+_{32},n^0_{33})$, 
$d_\delta\rightarrow {\bf{\varepsilon}}_1= (e^-_{11}, e^{0c}_{12},e^-_{13})$, 
$s_\delta\rightarrow {\bf{\varepsilon}}_2=(-e^-_{21}, e^{0c}_{22},e^-_{23})$ and
$b_\delta\rightarrow {\bf{\varepsilon}}_3= (-e^-_{31}, e^{0c}_{32},-e^-_{33})$;
 and the exclusion of the sum in the other expressions. 
Now, $-{\cal L}^{ll}_R$ is just ${\cal L}^{ll}_L$ with the substitutions 
$L\rightarrow R$  and $\{ e_{ij}, n_{ij} \}\rightarrow\{E_{ij},N_{ij}\}$ 
everywhere. 

If now one introduces instead of the mathematical leptons introduced in 
$\psi(\bar{6},1,1,6)\oplus\psi(1,6,\bar{6},1)$, the more natural set of lepton
fields ${\bf \it l} = (e,\mu,\tau)$, ${\bf \nu} =(\nu_e, \nu_\mu, \nu_\tau)$,
${\bf n} = (n_{12},n_{22},n_{32})$ and ${\bf e} = 
(e^0_{12},e^0_{22},e^0_{32})$, given by
\[
(e^-_{23},e^-_{11})_L=(n^-_{12},e^-)_L{\cal M};\quad
 (e^-_{13},e^-_{31})_L=(n^-_{22},\mu^-)_L{\cal M}; \quad
  (e^-_{33},e^-_{21})_L=(n^-_{32},\tau^-)_L {\cal M}\] 
\[(E^+_{23},E^+_{11})_L=(N^+_{12},e^+)_L{\cal M}; \quad 
  (E^+_{13},E^+_{31})_L=(N^+_{22},\mu^+)_L {\cal M};\quad
  (E^+_{33},E^+_{21})_L=(N^+_{32},\tau^+)_L{\cal M} \] 
\[(n^0_{23},n^0_{11})_L=(e^0_{12},\nu_e)_L {\cal M}; \quad
  (n^0_{13},n^0_{31})_L=(e^0_{22},\nu_\mu)_L {\cal M}; \quad
  (n^0_{33},n^0_{21})_L=(e^0_{32},\nu_\tau)_L{\cal M} \] 
\[(N^{0c}_{23},N^{0c}_{11})_L=(E^{0c}_{12},\nu_e^c)_L{\cal M}; \quad
  (N^{0c}_{13},N^{0c}_{31})_L=(E^{0c}_{22},\nu_\mu^c)_L {\cal M};\quad
  (N^{0c}_{33},N^{0c}_{21})_L=(E^{0c}_{32},\nu_\tau^c)_L {\cal M},\] 
where 
\[{\cal M}=\kappa\left(\begin{array}{cc}
Y_1 & Y_2 \\
Y_2 & -Y_1 \end{array}\right), \]
then the former Lagrangians can be put into the form
\begin{eqnarray}
{\cal L}_{CL}^{ql}&=&\frac{g}{\sqrt{2}}
\sum_{\delta=1}^3
\bigg[\kappa X^\mu_{\delta CL}\Big( Y_1 
(\bar\nu_e,-\bar\nu_\tau,-\bar\nu_\mu)_L\!\cdot\!\gamma_\mu U_{\delta L}+ 
Y_1(-\bar e^-,\bar\tau^-,\bar\mu^-)_L\!\cdot\!\gamma_\mu D_{\delta L}
\nonumber \\ &+& 
Y_2 (-\bar e^0_{12},\bar e^0_{32},\bar e^0_{22})_L\!\cdot\!\gamma_\mu 
U_{\delta L} + 
Y_2(\bar n^-_{12},-\bar n^-_{32},-\bar n^-_{22})_L\!\cdot\!\gamma_\mu 
D_{\delta L}\Big)\nonumber \\  &+&
Y^\mu_{\delta CL}\big(\bar{\bf n}_{L}^+\!\cdot\!\gamma_\mu U_{\delta L} + 
\bar{\bf e}_{L}^{c}\!\cdot\!\gamma_\mu D_{\delta L}\big)  \nonumber    \\ 
&+& \kappa Z^\mu_{\delta CL}\Big(
Y_1 (-\bar e^0_{22},-\bar e^0_{12},\bar e^0_{32})_L\!\cdot\!\gamma_\mu 
U_{\delta L} + 
Y_1 (\bar n^-_{22},\bar n^-_{12},-\bar n^-_{32})_L\!\cdot\!\gamma_\mu 
D_{\delta L} \nonumber \\  &+&
Y_2 (-\bar\nu_\mu,-\bar\nu_e,\bar\nu_\tau)_L\!\cdot\!\gamma_\mu U_{\delta L}+ 
Y_2 (\bar\mu^-,\bar e^-,-\bar\tau^-)_L\!\cdot\!\gamma_\mu D_{\delta L}
 \Big)+ h.c.\bigg] ,   \label{liqlr} 
\end{eqnarray}

\begin{eqnarray}
{\cal L}_{CL}^{ll}&=&\frac{g}{\sqrt{2}}
\bigg\{ D_{4CL}^\mu\, \kappa^2 \Big[Y_1^2 
\big(\bar{\bf\it l}^-_L\!\cdot\!\gamma_\mu{\bf\it l}^-_L + 
\bar{\bf\nu}_L\!\cdot\!\gamma_\mu{\bf\nu}_L\big) + 
Y_2^2\big( \bar{\bf n}^-_L\!\cdot\!\gamma_\mu{\bf n}^-_L + 
\bar{\bf e}_L\!\cdot\!\gamma_\mu{\bf e}_L \big) \nonumber \\ &-&
Y_1 Y_2 \big(\bar{\bf n}^-_L\!\cdot\!\gamma_\mu{\bf\it l}^-_L + 
\bar{\bf e}_L\!\cdot\!\gamma_\mu{\bf\nu}_L + h.c \big) \Big]  
+ D_{5CL}^\mu\Big[\bar{\bf n}_{L}^+\!\cdot\!\gamma_\mu {\bf n}_{L}^+  + 
\bar{\bf e}_{L}^{c}\!\cdot\!\gamma_\mu {\bf e}_{L}^{c}\Big] 
\nonumber \label {lillr}  \\ &+& 
D_{6CL}^\mu\, \kappa^2 \Big[
Y_1^2\big(\bar{\bf n}^-_L\!\cdot\!\gamma_\mu{\bf n}^-_L + 
\bar{\bf e}_L\!\cdot\!\gamma_\mu{\bf e}_L \big) +
Y_2^2\big(\bar{\bf\it l}^-_L\!\cdot\!\gamma_\mu{\bf\it l}^-_L + 
\bar{\bf\nu}_L\!\cdot\!\gamma_\mu{\bf\nu}_L\big) \nonumber \\ &+&
Y_1 Y_2 \big(\bar{\bf n}^-_L\!\cdot\!\gamma_\mu {\bf\it l}_L + 
\bar{\bf e}_L\!\cdot\!\gamma_\mu{\bf\nu}_L + h.c \big) \Big]  
+ P_{CL}^{0,\alpha}\, \kappa^2\Big[
Y_1^2\big(\bar{\bf\it l}_{\tau L}\!\cdot\!\gamma_\alpha{\bf\it l}_{1 L} - 
\bar{\bf\it l}_{e L}\!\cdot\!\gamma_\alpha{\bf\it l}_{2 L} -
\bar{\bf\it l}_{\mu L}\!\cdot\!\gamma_\alpha{\bf\it l}_{3L}\big) 
\nonumber \\ &+& 
Y_1Y_2\big(
\bar{\bf\it l}_{1 L}\!\cdot\!\gamma_\alpha{\bf\it l}_{2 L} +
\bar{\bf\it l}_{2 L}\!\cdot\!\gamma_\alpha{\bf\it l}_{3 L} -
\bar{\bf\it l}_{3 L}\!\cdot\!\gamma_\alpha{\bf\it l}_{1 L} +
\bar{\bf\it l}_{\tau L}\!\cdot\!\gamma_\alpha{\bf\it l}_{e L} -
\bar{\bf\it l}_{e L}\!\cdot\!\gamma_\alpha{\bf\it l}_{\mu L} -
\bar{\bf\it l}_{\mu L}\!\cdot\!\gamma_\alpha{\bf\it l}_{\tau L} \big) \nonumber
\\ &+& 
Y_2^2\big(
\bar{\bf\it l}_{1 L}\!\cdot\!\gamma_\alpha{\bf\it l}_{\mu L} +
\bar{\bf\it l}_{2 L}\!\cdot\!\gamma_\alpha{\bf\it l}_{\tau L} +
\bar{\bf\it l}_{3 L}\!\cdot\!\gamma_\alpha{\bf\it l}_{e L} \big) \Big]
+ P_{1CL}^{+,\alpha}\, \kappa \Big[ Y_1\big(
\bar{\bf\it l}^c_{3 L}\sigma_2\!\cdot\!\gamma_\alpha{\bf\it l}_{\mu L} +
\bar{\bf\it l}^c_{2 L}\sigma_2\!\cdot\!\gamma_\alpha{\bf\it l}_{\tau L}
\nonumber \\ &-&
\bar{\bf\it l}^c_{1 L}\sigma_2\!\cdot\!\gamma_\alpha{\bf\it l}_{e L} \big)
+ Y_2\big( \bar{ n}_{32}^+\gamma_\alpha { e}_{22}^0 +
\bar{ n}_{22}^+\gamma_\alpha { e}_{32}^0 - 
\bar{ n}_{12}^+\gamma_\alpha { e}_{12}^0 \big)  \Big]   \nonumber \\ &+&
P_{2CL}^{+,\alpha}\, \kappa \Big[Y_1\big(
\bar{ n}_{32}^+\gamma_\alpha { e}_{32}^0 -
\bar{ n}_{22}^+\gamma_\alpha { e}_{12}^0 - 
\bar{ n}_{12}^+\gamma_\alpha { e}_{22}^0 \big) + 
Y_2 \big(\bar{\bf\it l}^c_{1 L}\sigma_2\!\cdot\!\gamma_\alpha{\bf\it l}_{\mu L}+
\bar{\bf\it l}^c_{2 L}\sigma_2\!\cdot\!\gamma_\alpha{\bf\it l}_{e L} 
\nonumber \\ &-& 
\bar{\bf\it l}^c_{3 L}\sigma_2\!\cdot\!\gamma_\alpha{\bf\it l}_{\tau L} \big)
\Big] + h.c. \bigg\} ,   
\end{eqnarray}

\begin{eqnarray}
{\cal L}_{L}^{lW}&=&\frac{g}{\sqrt{6}}\bigg\{
\frac{W^0_{\mu L}}{\sqrt{2}}
\Big[ \bar{\bf n}^-_{L}\!\cdot\!\gamma^\mu {\bf n}^-_{L} 
-\bar{\bf e}_{L}\!\cdot\!\gamma^{\mu}{\bf e}_{L} +
\bar{\bf\it l}_{L}^-\!\cdot\!\gamma_\mu{\bf\it l}_{L}^- -
\bar{\bf\nu}_L\!\cdot\!\gamma_\mu{\bf\nu}_L \Big] \nonumber \\
&-& \Big[ W^+_{\mu L}\big(
\bar{\bf\nu}_L\!\cdot\!\gamma_\mu{\bf\it l}_{L}^- +
\bar{\bf e}_{L}\!\cdot\!\gamma^{\mu}{\bf n}^-_{L} \big) + h.c \Big] \bigg\},
\label{lilwr} 
\end{eqnarray}

\begin{eqnarray}
{\cal L}_{L}^{lH}&=&\frac{g}{\sqrt{2}}
\bigg\{\frac{H^{0\alpha}_{1L}}{\sqrt{2}}
\Big(\bar{\bf\it l}_{2 L}\!\cdot\!\gamma_\alpha{\bf\it l}_{1 L} +
\kappa^2 Y_1^2\big(\bar{\bf\it l}_{1L}\!\cdot\!\gamma_\alpha{\bf\it l}_{2L} -
\bar{\bf\it l}_{\tau L}\!\cdot\!\gamma_\alpha{\bf\it l}_{eL}\big) +
\kappa^2Y_2^2 \big(\bar{\bf\it l}_{eL}\!\cdot\!\gamma_\alpha{\bf\it l}_{\mu L} 
\nonumber \\ &-&
\bar{\bf\it l}_{3L}\!\cdot\!\gamma_\alpha{\bf\it l}_{1L}\big) 
 + \kappa^2 Y_1Y_2\big(
\bar{\bf\it l}_{3 L}\!\cdot\!\gamma_\alpha{\bf\it l}_{eL} +
\bar{\bf\it l}_{1 L}\!\cdot\!\gamma_\alpha{\bf\it l}_{\mu L} +
\bar{\bf\it l}_{e L}\!\cdot\!\gamma_\alpha{\bf\it l}_{2L} +
\bar{\bf\it l}_{\tau L}\!\cdot\!\gamma_\alpha{\bf\it l}_{1L}\big) \Big] 
\nonumber \\ &+& \label{lilhr}
H^{0\alpha}_{2L}\ \Big[ \bar e^{0c}_{32L}\gamma_\alpha E_{32L}^{0c} - 
\kappa^2 Y_1^2\big( \bar\mu^-_L\gamma_\alpha e^-_L + 
\bar n_{32L}^-\gamma_\alpha n^-_{22L}\big) + \kappa^2 Y_2^2\big(
\bar n_{22L}^-\gamma_\alpha n^-_{12L} \nonumber \\ &+&
 \bar\tau^-_L\gamma_\alpha \mu^-_L \big)
+ \kappa^2 Y_1Y_2\big(\bar n_{22L}^-\gamma_\alpha e^-_L \nonumber \\ &-&
\bar n_{32L}^-\gamma_\alpha \mu^-_L +\bar\mu^-_L\gamma_\alpha n^-_{12L} -
\bar\tau^-_L\gamma_\alpha n^-_{22L}\big)\Big] \nonumber \\
&+&\frac{H^{0\alpha}_{3L}}{\sqrt{2}}
\Big[\bar{\bf\it l}_{2 L}\sigma_3\!\cdot\!\gamma_\alpha{\bf\it l}_{1 L} +
\kappa^2 Y_1^2\big(
\bar{\bf\it l}_{\tau L}\sigma_3\!\cdot\!\gamma_\alpha{\bf\it l}_{eL} -
\bar{\bf\it l}_{1L}\sigma_3\!\cdot\!\gamma_\alpha{\bf\it l}_{2L} \big) +
\kappa^2Y_2^2 \big(
\bar{\bf\it l}_{3L}\sigma_3\!\cdot\!\gamma_\alpha{\bf\it l}_{1L}
\nonumber \\ &-&
\bar{\bf\it l}_{eL}\sigma_3\!\cdot\!\gamma_\alpha{\bf\it l}_{\mu L} \big) 
 - \kappa^2 Y_1Y_2\big(
\bar{\bf\it l}_{3 L}\sigma_3\!\cdot\!\gamma_\alpha{\bf\it l}_{eL} +
\bar{\bf\it l}_{1 L}\sigma_3\!\cdot\!\gamma_\alpha{\bf\it l}_{\mu L} +
\bar{\bf\it l}_{e L}\sigma_3\!\cdot\!\gamma_\alpha{\bf\it l}_{2L} +
\bar{\bf\it l}_{\tau L}\sigma_3\!\cdot\!\gamma_\alpha{\bf\it l}_{1L}\big) \Big]
\nonumber \\ &+& 
H^{0\alpha}_{4L}\big[  \bar n_{32L}^+\gamma_\alpha n_{12L}^+ - 
\kappa^2 Y_1^2\big( \bar e_{32L}^0\gamma_\alpha e_{22L}^0 + 
\bar\nu_{\mu L}\gamma_\alpha\nu_{eL}\big) - 
\kappa^2 Y_2^2\big(\bar e_{22L}^0\gamma_\alpha e_{12L}^0 \nonumber \\ &-&
\bar\nu_{\tau L}\gamma_\alpha\nu_{\mu L}\big)  +
\kappa^2 Y_2Y_1\big(\bar e_{22L}^0\gamma_\alpha\nu_{eL} - 
\bar e_{32L}^0\gamma_\alpha\nu_{\mu L} + \bar\nu_{\mu L}\gamma_\alpha e_{12L}^0
- \bar\nu_{\tau L}\gamma_\alpha e_{22L}^0\big) \Big]
\nonumber \\ &+&\frac{H^{0\alpha}_{5L}}{\sqrt{2}}\Big[
\bar{\bf\it l}_{3 L}^c\!\cdot\!\gamma_\alpha{\bf\it l}_{2 L}^c +
\kappa^2 Y_1^2\big(
\bar{\bf\it l}_{\mu L}\!\cdot\!\gamma_\alpha{\bf\it l}_{\tau L} -
\bar{\bf\it l}_{3L}\!\cdot\!\gamma_\alpha{\bf\it l}_{1L} \big) +
\kappa^2Y_2^2 \big(
\bar{\bf\it l}_{2L}\!\cdot\!\gamma_\alpha{\bf\it l}_{3L} \nonumber \\ &-&
\bar{\bf\it l}_{\tau L}\!\cdot\!\gamma_\alpha{\bf\it l}_{e L} \big) 
- \kappa^2 Y_1Y_2\big(
\bar{\bf\it l}_{3 L}\!\cdot\!\gamma_\alpha{\bf\it l}_{eL} +
\bar{\bf\it l}_{2 L}\!\cdot\!\gamma_\alpha{\bf\it l}_{\tau L} +
\bar{\bf\it l}_{\mu L}\!\cdot\!\gamma_\alpha{\bf\it l}_{3L} +
\bar{\bf\it l}_{\tau L}\!\cdot\!\gamma_\alpha{\bf\it l}_{1L}\big) \Big]
\nonumber \\ &+& \frac{H^{0\alpha}_{6L}}{\sqrt{2}}\Big[
\bar{\bf\it l}_{3 L}^c\sigma_3\!\cdot\!\gamma_\alpha{\bf\it l}_{2 L}^c +
\kappa^2 Y_1^2\big(
\bar{\bf\it l}_{3L}\sigma_3\!\cdot\!\gamma_\alpha{\bf\it l}_{1L} -
\bar{\bf\it l}_{\mu L}\sigma_3\!\cdot\!\gamma_\alpha{\bf\it l}_{\tau L} \big) 
 + \kappa^2Y_2^2 \big(
 \bar{\bf\it l}_{\tau L}\sigma_3\!\cdot\!\gamma_\alpha{\bf\it l}_{e L}
 \nonumber \\ &-&
\bar{\bf\it l}_{2L}\sigma_3\!\cdot\!\gamma_\alpha{\bf\it l}_{3L} \big)  
+ \kappa^2 Y_1Y_2\big(
\bar{\bf\it l}_{3 L}\sigma_3\!\cdot\!\gamma_\alpha{\bf\it l}_{eL} +
\bar{\bf\it l}_{2 L}\sigma_3\!\cdot\!\gamma_\alpha{\bf\it l}_{\tau L} +
\bar{\bf\it l}_{\mu L}\sigma_3\!\cdot\!\gamma_\alpha{\bf\it l}_{3L} +
\bar{\bf\it l}_{\tau L}\sigma_3\!\cdot\!\gamma_\alpha{\bf\it l}_{1L}\big) \Big]
\nonumber \\ & +& h.c.\bigg\},
\end{eqnarray}

\begin{eqnarray}
{\cal L}_{L}^{lA}&=&\frac{g}{2\sqrt{2}}\bigg\{A^\alpha_{1HL}\Big[
\bar{\bf\it l}_{1 L}^c\!\cdot\!\gamma_\alpha{\bf\it l}_{1 L}^c -
\bar{\bf\it l}_{3 L}^c\!\cdot\!\gamma_\alpha{\bf\it l}_{3 L}^c  + 
\kappa^2Y_1^2\big(
\bar{\bf\it l}_{e L}\!\cdot\!\gamma_\alpha{\bf\it l}_{e L} -
\bar{\bf\it l}_{3 L}\!\cdot\!\gamma_\alpha{\bf\it l}_{3 L} \big) 
\nonumber \\ &+& 
\kappa^2Y_2^2\big(
\bar{\bf\it l}_{1 L}\!\cdot\!\gamma_\alpha{\bf\it l}_{1 L} -
\bar{\bf\it l}_{\tau L}\!\cdot\!\gamma_\alpha{\bf\it l}_{\tau L} \big) +
\kappa^2\left(Y_1^2 - Y_2^2\right)\big(
\bar{\bf\it l}_{2 L}\!\cdot\!\gamma_\alpha{\bf\it l}_{2 L} -
\bar{\bf\it l}_{\mu L}\!\cdot\!\gamma_\alpha{\bf\it l}_{\mu L} \big)
\nonumber \\ &+&
\kappa^2Y_1Y_2\big(
2  \bar{\bf\it l}_{2 L}\!\cdot\!\gamma_\alpha{\bf\it l}_{\mu L} -
\bar{\bf\it l}_{1 L}\!\cdot\!\gamma_\alpha{\bf\it l}_{e L} -
\bar{\bf\it l}_{3 L}\!\cdot\!\gamma_\alpha{\bf\it l}_{\tau L} + h.c.\big)\Big]
\nonumber \\ &+&
\frac{A^\alpha_{2HL}}{\sqrt{3}}\Big[\ 
\bar{\bf\it l}_{1 L}^c\!\cdot\!\gamma_\alpha{\bf\it l}_{1 L}^c -
2 \bar{\bf\it l}_{2 L}^c\!\cdot\!\gamma_\alpha{\bf\it l}_{2 L}^c +
\bar{\bf\it l}_{3 L}^c\!\cdot\!\gamma_\alpha{\bf\it l}_{3 L}^c + 
\bar{\bf\it l}_{\mu L}\!\cdot\!\gamma_\alpha{\bf\it l}_{\mu L} +
\bar{\bf\it l}_{2 L}\!\cdot\!\gamma_\alpha{\bf\it l}_{2 L} 
\nonumber \\ &+&
\kappa^2\left(Y_1^2 - 2 Y_2^2\right)\big(
\bar{\bf\it l}_{e L}\!\cdot\!\gamma_\alpha{\bf\it l}_{e L} +
\bar{\bf\it l}_{3 L}\!\cdot\!\gamma_\alpha{\bf\it l}_{3 L} \big) 
- \kappa^2\left(2 Y_1^2 - Y_2^2\right)\big(
\bar{\bf\it l}_{1 L}\!\cdot\!\gamma_\alpha{\bf\it l}_{1 L} +
\bar{\bf\it l}_{\tau L}\!\cdot\!\gamma_\alpha{\bf\it l}_{\tau L}\big)
\nonumber \\ &+&
3\kappa^2 Y_1 Y_2\big(
\bar{\bf\it l}_{3 L}\!\cdot\!\gamma_\alpha{\bf\it l}_{\tau L} - 
\bar{\bf\it l}_{1 L}\!\cdot\!\gamma_\alpha{\bf\it l}_{e L} + h.c.\big)\Big]
\nonumber \\ &+&
\frac{A^\alpha_{2AL}}{\sqrt{3}}\Big[
2 \bar{\bf\it l}_{2 L}^c\sigma_3\!\cdot\!\gamma_\alpha{\bf\it l}_{2 L}^c - 
\bar{\bf\it l}_{1L}^c\sigma_3\!\cdot\!\gamma_\alpha{\bf\it l}_{1 L}^c -
\bar{\bf\it l}_{3L}\sigma_3\!\cdot\!\gamma_\alpha{\bf\it l}_{3 L} +
 \bar{\bf\it l}_{2 L}\sigma_3\!\cdot\!\gamma_\alpha{\bf\it l}_{2 L} +
\bar{\bf\it l}_{\mu L}\sigma_3\!\cdot\!\gamma_\alpha{\bf\it l}_{\mu L} 
\nonumber \\ &+&
\kappa^2\left(Y_1^2 - 2 Y_2^2\right)\big(
\bar{\bf\it l}_{e L}\sigma_3\!\cdot\!\gamma_\alpha{\bf\it l}_{e L} +
\bar{\bf\it l}_{3 L}\sigma_3\!\cdot\!\gamma_\alpha{\bf\it l}_{3 L} \big) 
\nonumber \\ &-&
 \kappa^2\left(2 Y_1^2 - Y_2^2\right)\big(
\bar{\bf\it l}_{1 L}\sigma_3\!\cdot\!\gamma_\alpha{\bf\it l}_{1 L} +
\bar{\bf\it l}_{\tau L}\sigma_3\!\cdot\!\gamma_\alpha{\bf\it l}_{\tau L}\big)
\nonumber \\ &+&
3\kappa^2 Y_1 Y_2\big(
\bar{\bf\it l}_{3 L}\sigma_3\!\cdot\!\gamma_\alpha{\bf\it l}_{\tau L} - 
\bar{\bf\it l}_{1 L}\sigma_3\!\cdot\!\gamma_\alpha{\bf\it l}_{e L} + 
h.c.\big)\Big]
\nonumber \\ &+& 
A^\alpha_{1AL}\big[
\bar{\bf\it l}_{3L}^c\sigma_3\!\cdot\!\gamma_\alpha{\bf\it l}_{3 L}^c -
\bar{\bf\it l}_{1 L}^c\sigma_3\!\cdot\!\gamma_\alpha{\bf\it l}_{1 L}^c +
\kappa^2Y_1^2\big(
\bar{\bf\it l}_{e L}\sigma_3\!\cdot\!\gamma_\alpha{\bf\it l}_{e L} -
\bar{\bf\it l}_{3 L}\sigma_3\!\cdot\!\gamma_\alpha{\bf\it l}_{3 L} \big) 
\nonumber \\ &+& 
\kappa^2Y_2^2\big(
\bar{\bf\it l}_{1 L}\sigma_3\!\cdot\!\gamma_\alpha{\bf\it l}_{1 L} -
\bar{\bf\it l}_{\tau L}\sigma_3\!\cdot\!\gamma_\alpha{\bf\it l}_{\tau L} \big) +
\kappa^2\left(Y_1^2 - Y_2^2\right)\big(
\bar{\bf\it l}_{2 L}\sigma_3\!\cdot\!\gamma_\alpha{\bf\it l}_{2 L} -
\bar{\bf\it l}_{\mu L}\sigma_3\!\cdot\!\gamma_\alpha{\bf\it l}_{\mu L} \big)
\nonumber \\ &+&
\kappa^2Y_1Y_2\big(
2  \bar{\bf\it l}_{2 L}\sigma_3\!\cdot\!\gamma_\alpha{\bf\it l}_{\mu L} -
\bar{\bf\it l}_{1 L}\sigma_3\!\cdot\!\gamma_\alpha{\bf\it l}_{e L} -
\bar{\bf\it l}_{3 L}\sigma_3\!\cdot\!\gamma_\alpha{\bf\it l}_{\tau L} 
 + h.c.\big)\Big] \bigg\},
\end{eqnarray} 

\begin{eqnarray}
{\cal L}_{L}^{lB}&=&\frac{g}{\sqrt{2}}
\bigg\{\frac{B^{-\alpha}_{1L}}{\sqrt{2}}\Big[
\bar e^{0c}_{12L}\gamma_\alpha n_{12L}^+ - 
\bar e^{0c}_{32L}\gamma_\alpha n_{32L}^+ +
\kappa^2\left(Y_1^2 - Y_2^2\right)\big( 
\bar \mu^-_L\gamma_\alpha \nu_{\mu L} -
\bar n_{22L}^- \gamma_\alpha e^{0}_{22L}\big) 
 \nonumber \\ &+&  
 \kappa^2 Y_1^2 \big(\bar n_{32L}^-\gamma_\alpha e^{0}_{32L} - 
 \bar e^-_L\gamma_\alpha\nu_{eL}\big)+  
 \kappa^2 Y_2^2 \big( \bar \tau^-_L\gamma_\alpha\nu_{\tau L} - 
\bar n_{12L}^-\gamma_\alpha e^{0}_{12L} \big) +
\kappa^2Y_1Y_2\big(\bar e^-_L\gamma_\alpha e^{0}_{12L} 
 \nonumber \\ &+&  
\bar n_{12L}^-\gamma_\alpha\nu_{eL} +\bar n_{32L}^-\gamma_\alpha\nu_{\tau L} 
-2 \bar \mu^-_L\gamma_\alpha e^{0}_{22L} - 
2 \bar n_{22L}^- \gamma_\alpha\nu_{\mu L}  + 
\bar \tau^-_L\gamma_\alpha e^{0}_{32L} \big) \Big] 
\nonumber \\ &+&
B^{-\alpha}_{2L}\ \Big[
\bar e^{0c}_{12L}\gamma_\alpha n_{22L}^+ +
\kappa^2 Y_1^2 \big(\bar e^-_L\gamma_\alpha\nu_{\tau L} -
\bar n_{22L}^-\gamma_\alpha e^{0}_{12L} \big) +
\kappa^2 Y_2^2 \big(\bar n_{12L}^-\gamma_\alpha e^{0}_{32L}  -
\bar \mu^-_L\gamma_\alpha\nu_{e L}  \big)
\nonumber \\ &-&
\kappa^2Y_1Y_2\big(\bar e^-_L\gamma_\alpha e^{0}_{32L} +
\bar \mu^-_L\gamma_\alpha e^{0}_{12L} +
\bar n_{22L}^-\gamma_\alpha\nu_{eL} +
\bar n_{12L}^-\gamma_\alpha\nu_{\tau L}\big) \Big] 
\nonumber \\ &+&
B^{-\alpha}_{3L}\Big[
\bar e^{0c}_{12L}\gamma_\alpha n_{32L}^+ +
\kappa^2 Y_1^2 \big(\bar e^-_L\gamma_\alpha\nu_{\mu L} +
\bar n_{22L}^-\gamma_\alpha e^{0}_{32L} \big) +
\kappa^2 Y_2^2 \big(\bar n_{12L}^-\gamma_\alpha e^{0}_{22L}  +
\bar \mu^-_L\gamma_\alpha\nu_{\tau L}  \big)
\nonumber \\ &-&
\kappa^2Y_1Y_2\big(\bar e^-_L\gamma_\alpha e^{0}_{22L} -
\bar \mu^-_L\gamma_\alpha e^{0}_{22L} -
\bar n_{22L}^-\gamma_\alpha\nu_{\tau L} +
\bar n_{12L}^-\gamma_\alpha\nu_{\mu L}\big) \Big] 
\nonumber \\ &+&
 B^{-\alpha}_{4L}\ \Big[
\bar e^{0c}_{22L}\gamma_\alpha n_{12L}^+ +
\kappa^2 Y_1^2 \big(\bar \tau^-_L\gamma_\alpha\nu_{e L} -
\bar n_{12L}^-\gamma_\alpha e^{0}_{22L} \big) +
\kappa^2 Y_2^2 \big(\bar n_{32L}^-\gamma_\alpha e^{0}_{12L}  -
\bar e^-_L\gamma_\alpha\nu_{\mu L}  \big)
\nonumber \\ &-&
\kappa^2Y_1Y_2\big(\bar e^-_L\gamma_\alpha e^{0}_{22L} +
\bar \tau^-_L\gamma_\alpha e^{0}_{12L} +
\bar n_{32L}^-\gamma_\alpha\nu_{e L} +
\bar n_{12L}^-\gamma_\alpha\nu_{\mu L}\big) \Big] 
\nonumber \\ &+& 
B^{-\mu}_{5L}\ \Big[
\bar e^{0c}_{32L}\gamma_\alpha n_{12L}^+ +
\kappa^2 Y_1^2 \big(\bar \mu^-_L\gamma_\alpha\nu_{e L} +
\bar n_{32L}^-\gamma_\alpha e^{0}_{22L} \big) +
\kappa^2 Y_2^2 \big(\bar n_{22L}^-\gamma_\alpha e^{0}_{12L}  +
\bar \tau^-_L\gamma_\alpha\nu_{\mu L}  \big)
\nonumber \\ &-&
\kappa^2Y_1Y_2\big(\bar \mu^-_L\gamma_\alpha e^{0}_{12L} -
\bar \tau^-_L\gamma_\alpha e^{0}_{22L} -
\bar n_{32L}^-\gamma_\alpha\nu_{\mu L} +
\bar n_{22L}^-\gamma_\alpha\nu_{e L}\big) \Big] 
\nonumber \\ &+& 
\frac{B^{-\mu}_{6L}}{\sqrt{6}}\Big[
\bar e^{0c}_{12L}\gamma_\alpha n_{12L}^+ - 
2\bar e^{0c}_{22L}\gamma_\alpha n_{22L}^+  + 
\bar e^{0c}_{32L}\gamma_\alpha n_{32L}^+  -
 \bar \mu^-_L\gamma_\alpha \nu_{\mu L} -
\bar n_{22L}^- \gamma_\alpha e^{0}_{22L}  
\nonumber \\ &+&
\kappa^2\left(2 Y_1^2 - Y_2^2\right)\big( 
\bar n_{12L}^-\gamma_\alpha e^{0}_{12L}  +
\bar \tau^-_L\gamma_\alpha\nu_{\tau L}\big) -
\kappa^2\left( Y_1^2 - 2 Y_2^2\right)\big( 
\bar n_{32L}^-\gamma_\alpha e^{0}_{32L} +
 \bar e^-_L\gamma_\alpha\nu_{eL}\big) 
\nonumber \\ &+&
3 \kappa^2Y_1Y_2\big( \bar e^-_L\gamma_\alpha e^{0}_{12L} +
\bar n_{12L}^-\gamma_\alpha\nu_{eL} - \bar n_{32L}^-\gamma_\alpha\nu_{\tau L} 
- \bar \tau^-_L\gamma_\alpha e^{0}_{32L} \big) \Big] 
\nonumber \\ &+&
B^{-\mu}_{7L}\Big[
\bar e^{0c}_{22L}\gamma_\alpha n_{32L}^+ +
\kappa^2 Y_1^2 \big(\bar n_{12L}^-\gamma_\alpha e^{0}_{32L} -
\bar \tau^-_L\gamma_\alpha\nu_{\mu L} \big) +
\kappa^2 Y_2^2 \big(\bar e^-_L\gamma_\alpha\nu_{\tau L}  - 
\bar n_{32L}^-\gamma_\alpha e^{0}_{22L} \big)
\nonumber \\ &-&
\kappa^2Y_1Y_2\big(\bar e^-_L\gamma_\alpha e^{0}_{32L} +
\bar \tau^-_L\gamma_\alpha e^{0}_{22L} +
\bar n_{32L}^-\gamma_\alpha\nu_{\mu L} +
\bar n_{12L}^-\gamma_\alpha\nu_{\tau L}\big) \Big] 
\nonumber \\ &+&
B^{-\mu}_{8L}\ \Big[
\bar e^{0c}_{32L}\gamma_\alpha n_{22L}^+ +
\kappa^2 Y_1^2 \big(\bar n_{32L}^-\gamma_\alpha e^{0}_{12L} -
\bar \mu^-_L\gamma_\alpha\nu_{\tau L} \big) +
\kappa^2 Y_2^2 \big(\bar \tau^-_L\gamma_\alpha\nu_{e L}  - 
\bar n_{22L}^-\gamma_\alpha e^{0}_{32L} \big)
\nonumber \\ &-&
\kappa^2Y_1Y_2\big(\bar \mu^-_L\gamma_\alpha e^{0}_{32L} +
\bar \tau^-_L\gamma_\alpha e^{0}_{12L} +
\bar n_{32L}^-\gamma_\alpha\nu_{e L} +
\bar n_{22L}^-\gamma_\alpha\nu_{\tau L}\big) \Big] 
\nonumber \\ &+&
 h.c.\bigg].
\end{eqnarray}
where we have used the following leptonic doublets: ${\bf\it l}_l \equiv
(l^-,\nu_l)$ for $l = e, \mu, \tau$ and ${\bf\it l}_i \equiv
(n_{i2}^-,e^0_{i2})$ for $i=1,2,3$; and the rotating matrices
\begin{equation}
\sigma_2 = \left( \begin{array}{c c} 0 & -1\\ 1&0 \end{array} \right) ;
\qquad 
\sigma_3 = \left( \begin{array}{c c} 1 & 0\\ 0 & -1 \end{array} \right).
\end{equation}

\section{\uno Stability of the Proton}
In the subspace of the fundamental representation of 
SU(6)$_{CR}\otimes$SU(6)$_{CL}$ the baryon number for G can be 
associated with the $12\times 12$ diagonal submatrix\\ 
B=$Diag.[(1/3,1/3,1/3,0,0,0)\oplus (1/3,1/3,1/3,0,0,0)]$. Since this 
matrix does not correspond to a lineal combination of generators in  G 
then the baryon number is not gauged in this model (there is not a gauge bosson
in $G$ associated with B). 

Now due to the stated directions of the VEVs for $\phi_i, i=1-4$ in section  
2.3, it is a matter of algebra to show that B$\langle\phi_i\rangle =0, i=1-4$. 
Therefore B is not broken 
spontaneously by the 
set of Higgs fields used for the breaking of G down to 
SU(3)$_C\otimes$U(1)$_{EM}$. So, B is perturbatively conserved in the context
of the model presented here, and the proton remains perturbatively stable. 

Another way to see this is to use 
t'Hooft\cite{toof} argument and to consider two generators $BL$ and $\Theta$ 
in the subspace of the fundamental representation 
for SU(6)$_{CR}\otimes$SU(6)$_{CL}$ defined as 
\[
BL=Diag.[(1,1,1,-1,-1,-1)\oplus (1,1,1,-1,-1,-1)]
\]
which is a generator of the G algebra which distinguishes baryon and 
lepton number, and
\[
\Theta = Diag.[(1,1,1,1,1,1)\oplus(1,1,1,1,1,1)]
\] 
which generates a U(1)$_{\Theta}$ global symmetry of the model. 
$BL$ and $\Theta$ are spontaneously 
broken by $\langle\phi_i\rangle,i=1-4$, but the lineal combination 
$B=(BL+\Theta$)/6 is not.

\section{\uno Concluding remarks}
We have studied in detail various aspects of the $[SU(6)]^4\times Z_4$
grand-unification model, using the fields in the representation
$\psi(144)=Z_4\psi(\bar{6},1,1,6)$ as presented in the main text. The most
outstanding features of the model are:
\begin{itemize}
\item The evolution from low to high energies of the gauge couplings in $G$, 
meet together at a single point at the scale $M\sim 10 ^8$
GeV, in good agreement with precision data tests of the SM. We emphasize
that this is the only realistic (small number of low energy Higgs
doublets) non supersymmetric model for three families which descends to the
SM group in one single step, as a detailed analysis shows~\cite{nonsusy}.
\item The low unification scale does not conflict with data on proton
stability because baryon number is perturbatively conserved. 
\item Unlike the model presented in Ref.~\cite{elias}, our $\psi(144)$ does
not contain mirror fermions, and it is not vectorlike
with respect to $G$. Therefore the survival hypothesis~\cite{sh} and the
decoupling theorem~\cite{ac} can be properly implemented, in such a way that
all the exotic fields in $\psi(144)$ get very large masses (of the order
of the unification scale). 
\item At tree level the only ordinary charged fermion field which get mass
(of the order $M_Z$) is the $t$ quark, in consistence with the modified
horizontal survival hypothesis~\cite{ponce}. Masses for the other standard
charged fermion fields should be generated as radiative corrections. 
\item The mass terms for the neutral particles of the model show that a
generational (three family) see-saw mechanism may easily  be implement in
order to explain the small neutrino masses~\cite{pzab}. 
\end{itemize}
\section{\uno Acknowledgments}
This work was partially supported by CONACyT in Mexico, and COLCIENCIAS and
Banco de la Rep\'ublica in Colombia.


\begin{appendix}
\begin{center}
{\bf APPENDIX A}
\end{center}
The terminology used in the main text has been properly translated from 
classical papers on grand unified theories from fifteen years ago. 

{\bf Survival Hypothesis}\cite{sh}. For a symmetry group $G$ with 
$G^\prime\subset G$, if $G$ is spontaneously broken down to $G^\prime$ 
at the mass scale $M$ 
$(G\stackrel{M}{\longrightarrow}G^\prime)$, then according to the survival 
hypothesis, any set of fermion fields which are  vector representations
of $G^\prime$ should get masses of order $M$. In other words, ``at each energy
scale the only relevant fermion are those which are chiral with
respect to the surviving symmetry''.

{\bf Extended Survival Hypothesis}\cite{aguila}. 
It claims that only the scalar fields which acquire VEVs at a particular mass 
scale, acquire masses at that scale, with the rest of the 
scalar fields acquiring masses at the unification scale. In other words,
``Higgses acquire the maximum mass compatible with the pattern of symmetry
breaking"

{\bf Horizontal Survival Hypothesis}\cite{hsh}.
It claims that only the particles in the heaviest family of 
quarks and leptons acquire masses at tree level from dimension four Yukawa 
couplings, with all the other families getting masses via radiative 
corrections.

{\bf Modified Horizontal Survival Hypothesis}\cite{ponce}. 
It claims that for a universe with three families, only the top quark and 
$\nu_\tau$ acquire tree level masses (the last one lower down with the 
appropriate see-saw mechanism), with the masses for all the other known 
fermions generated via radiative corrections.

\end{appendix}


\vspace{2cm}

\begin{appendix}
\begin{center}
APPENDIX B
\end{center}
In this appendix we introduce some mathematical definitions used in the main 
text.

First, the diagonal entries in Eq.(\ref{3}) are related to the physical fields
by  
\begin{equation}
\left(\begin{array}{c}
A_1 \\ A_2 \\ A_3 \\ A_4 \\ A_5 \\ A_6 \end{array} \right)_{L(R)} = 
\left( \begin{array}{ccccc}
1/\sqrt{6}  & 1/2  & 1/\sqrt{12}  & 1/2  & 1/\sqrt{12}  \\
-1/\sqrt{6} & 1/2  & 1/\sqrt{12}  &- 1/2 & -1/\sqrt{12} \\  
1/\sqrt{6}  & 0    & -2/\sqrt{12} & 0    & -2/\sqrt{12} \\  
-1/\sqrt{6} & 0    & -2/\sqrt{12} & 0    & 2/\sqrt{12}  \\  
1/\sqrt{6}  & -1/2 & 1/\sqrt{12}  & -1/2 & 1/\sqrt{12}  \\  
-1/\sqrt{6} & -1/2 & 1/\sqrt{12}  & 1/2  & -1/\sqrt{12} \end{array} \right)
\left(\begin{array}{c}
W^0 \\ A_{1H} \\ A_{2H} \\ A_{1A} \\ A_{2A} \end{array}\right)_{L(R)}  
\end{equation}
\noindent
where the gauge fields $W^0_{L(R)},A_{1HL(R)},A_{2HL(R)},A_{1AL(R)}$, 
and $A_{2AL(R)}$ are related to the following set of diagonal 
generators of SU(6)$_{L(R)}$:\\
$Y_{WL(R)}=Diag(1,-1,1,-1,1,-1)/\sqrt{3}; 
Y_{A_{1HL(R)}}=Diag(1,1,0,0,-1,-1)/\sqrt{2};\\
Y_{A_{2HL(R)}}=Diag(1,1,-2,-2,1,1)/\sqrt{6}; 
Y_{A_{1AL(R)}}=Diag(1,-1,0,0,-1,1)/\sqrt{2}$,\\ 
and $Y_{A_{2AL(R)}}=Diag(1,-1,-2,2,1,-1)/\sqrt{6}$, respectively.

The primed fields in Eq.(\ref{3}) $B^{\prime\pm}_l,l=1,6,9$ are related to 
a set of unprimed ones by the equations
\begin{equation}
\left(\begin{array}{c}
W^\pm_{L(R)} \\ B^\pm_{1L(R)} \\ B^\pm_{6L(R)} \end{array} \right) = \left(
\begin{array}{ccc}
1/\sqrt{3} & 1/\sqrt{3} & 1/\sqrt{3} \\
1/\sqrt{2} & 0 & -1/\sqrt{2} \\
1/\sqrt{6} & -2/\sqrt{6} & 1/\sqrt{6} \end{array} \right)\left(\begin{array}{c}
B^{\prime\pm}_1 \\ B^{\prime\pm}_{6} \\ B^{\prime\pm}_9 \end{array} 
\right)_{L(R)}.
\end{equation}
In order to simplify matters we have defined
\begin{equation}
\left(\begin{array}{c}
H^0_1 \\ H^0_3 \\ H^0_5 \\ H^0_6 \end{array} \right)_{L(R)} = 
\frac{1}{\sqrt{2}}\left(\begin{array}{cccc}
 1 & 1 &  0 & 0 \\
-1 & 1 &  0 & 0 \\
 0 & 0 &  1 & 1 \\
 0 & 0 & -1 & 1  \end{array} \right) 
\left(\begin{array}{c}
H^{\prime 0}_1 \\ H^{\prime 0}_3 \\ H^{\prime 0}_5 \\ H^{\prime 0}_6 
\end{array} \right)_{L(R)}.
\end{equation}

With the former definitions $W^0_{L(R)}$ and $W^\pm_{L(R)}$ are the gauge 
fields associated with the gauge group SU(2)$_{L(R)}$of the left-right 
symmetric model.
\end{appendix}


\end{document}